\documentclass[a4paper,12pt]{article}
\usepackage{amssymb,amsmath,amsthm}
\usepackage{a4wide}
\usepackage{pgf,tikz}
\usetikzlibrary{arrows}
\usepackage{epsfig}
\usepackage{color}

\usepackage{hyperref}

\usepackage{authblk}

\usepackage{subcaption}

\newcommand{\beq}{\begin{equation}}  
\newcommand{\eeq}{\end{equation}}  
\newcommand{\bea}{\begin{eqnarray}} 
\newcommand{\eea}{\end{eqnarray}}   
\newcommand{\bear}{\begin{array}}  
\newcommand{\eear}{\end{array}}

\newtheorem{thm}{Theorem}[section]

\newtheorem{lem}[thm]{Lemma}

\newenvironment{prf}{\trivlist \item [\hskip 
\labelsep {\bf Proof:}]\ignorespaces}{\qed \endtrivlist}

\theoremstyle{definition}

\newtheorem{remark}[thm]{Remark}

\newcommand{\F}{{\mathbb F}}

\newcommand{\Q}{{\mathbb Q}}
\newcommand{\Z}{{\mathbb Z}}
\newcommand{\C}{{\mathbb C}}
\newcommand{\R}{{\mathbb R}}

\newcommand{\rd}{\mathrm{d}}

\newcommand\la{{\lambda}}

\newcommand\al{{\alpha}}
\newcommand\be{{\beta}}

\newcommand\gam{{\gamma}}

\newcommand\om{{\omega}}

\newcommand\tmu{\tilde{\mu}}
\newcommand\si{{\sigma}}

\newcommand\tx{\tilde{x}}

\newcommand\eps{{\epsilon}}

\def\lc{\left\lfloor}   
\def\rc{\right\rfloor}

\begin{document}


\title{Deformations of cluster mutations and invariant presymplectic forms}
\author[1]{Andrew N. W. Hone} 
\author[2]{Theodoros E. Kouloukas} %
\affil[1]{School of Mathematics, Statistics \&  Actuarial Science, 
University of Kent, 
Canterbury CT2 7FS, U.K. 
}
\affil[2]{School of Mathematics and Physics, University of Lincoln,
Lincoln LN6 7TS, U.K.}

\maketitle

\begin{abstract} 
We consider deformations of sequences of cluster mutations in finite type cluster algebras, which destroy the Laurent 
property but preserve the presymplectic structure defined by the exchange matrix. The simplest example is the Lyness 5-cycle, 
arising from the cluster algebra of type $A_2$: this deforms to the Lyness family of integrable symplectic maps in the plane. For 
types $A_3$ and $A_4$ we find suitable conditions such that the deformation produces a two-parameter family of 
Liouville integrable maps (in dimensions two and four, respectively). We also perform Laurentification for these maps, 
by lifting them to a higher-dimensional space of tau functions with a cluster algebra structure, 
where the Laurent property is restored. More general types of deformed mutations 
associated with affine Dynkin quivers 
are shown to correspond to four-dimensional symplectic maps arising as reductions of the discrete sine-Gordon equation.  
\end{abstract}

\section{Lyness maps and Zamolodchikov periodicity} 

\setcounter{equation}{0}

It was observed by Lyness in 1942 \cite{lyness} that the recurrence 
\beq\label{l5}
x_{n+2}x_n=x_{n+1}+1 
\eeq 
generates the sequence 
\beq\label{5cycle}
x_0,x_1,\frac{x_1+1}{x_0},\frac{x_0+x_1+1}{x_0x_1}, 
  \frac{x_0+1}{x_1}, x_0,x_1,\ldots , 
\eeq 
 which repeats  with period five. 
The Lyness 5-cycle also arises in Coxeter's  
frieze patterns \cite{coxeter}, or as a simple example 
of Zamolodchikov periodicity in integrable quantum field theories \cite{zam}, 
which can be understood in terms of the associahedron $K_4$ and the cluster 
algebra defined by the $A_2$ Dynkin quiver \cite{fz2}, and this leads to 
a connection with Abel's pentagon identity for the dilogarithm \cite{nak}. 
The birational map of the plane corresponding to the recurrence (\ref{l5}), that is  
\beq\label{ly5} 
(x,y)\mapsto \left(y, \frac{y+1}{x}\right), 
\eeq 
also appears in the theory of   the Cremona group: 
as conjectured by Usnich and proved by Blanc \cite{blanc}, the birational transformations 
of the plane that preserve the symplectic form 
\beq\label{omegaplane} 
\om = \frac{1}{xy}\, \rd x \wedge \rd y, 
\eeq 
are generated by $SL(2,\Z)$, the torus and 
transformation (\ref{ly5}). 

More generally, the birational map 
\beq\label{ly} 
\varphi: \quad (x,y)\mapsto \left(y, \frac{ay+b}{x}\right), 
\eeq 
with two parameters $a,b$ is also  referred to as the  Lyness map. 
By rescaling $(x,y)\to (ax,ay)$, 
the parameter $a\neq 0$ 
can be removed, 
so that 
this is really a one-parameter family, which is described in \cite{eschr} 
as ``the simplest singular map of the plane.'' 
There are 
also analogous  recurrences  in higher dimensions, 
given by the family 
$$
x_{n+N}x_n=\sum_{j=1}^{N-1}x_{n+j} +b, 
$$ 
which have been shown to admit $\lc\frac{N}{2}\rc$ independent first integrals for each order $N$  \cite{tran}.

Unlike the special case $b=a^2$, which can be rescaled  to (\ref{ly5}), 
in general the orbits of (\ref{ly}) do not all have the same period, and  generic orbits are not periodic 
over an infinite field (e.g.\ $\Q,\R$ or $\mathbb C$). 
Moreover, while the iterates in (\ref{5cycle}) are Laurent polynomials in the initial values $x_0,x_1$ 
with integer coefficients, which is one of the characteristic features of the cluster variables in a 
cluster algebra, the iterates of (\ref{ly}) are not Laurent polynomials unless $b=a^2$. 
However, the general map does preserve the same symplectic     
form (\ref{omegaplane}), and there is a conserved quantity 
$K=K(x,y)$ given by 
\beq\label{integral}
K=\frac{ xy(x+y)+a(x^2+y^2) +(a^2+b)(x+y)+ab}{xy}. 
\eeq 

Thus the Lyness map (\ref{ly}) is integrable in the Liouville sense, and can be considered as a deformation 
of the periodic map (\ref{ly5}) which arises from mutations in a finite type cluster algebra.  The purpose of this work is 
to  consider how other integrable maps can be obtained from deformations of cluster mutations. The Zamolodchikov periodicity of 
Y-systems or T-systems associated with finite type root systems has been extended and generalized in various ways (see \cite{gal, kun, pyl} 
and references), but as far as we are aware the deformations we consider are new.

Following the framework of 
cluster algebras, we start from  a  quiver $Q$ (without 1- or 2-cycles) associated with a skew-symmetric {\it{exchange matrix}} $B=(b_{ij}) \in \mathrm{Mat}_N(\mathbb{Z})$ and  an
$N$-tuple of \textit{cluster variables} ${\bf x} = (x_1,x_2,\ldots, x_N)$. Here we consider the cluster variables $x_i$ taking values in a field  $\mathbb{F}$; the main cases
of interest are  
$\F=\R$ or $\C$, but for some of our later analysis it will be convenient to consider $x_i\in\Q\subset\Q_p$. 
The initial seed is denoted $(B, {\bf x})$.  
Now,  for each integer
$k\in [1,N]$ we define a  mutation ${\mu}_k$ which produces a new seed
$(B', {\bf x}')={\mu}_k (B, {\bf x})$, where  $B'=(b_{ij}')$ with
\beq\label{matmut}
b_{ij}' =  \begin{cases}
 -b_{ij} &\text{if}  \,\,i=k \,\, \text{or} \,\, j=k , \\
 b_{ij}+\text{sgn}(b_{ik})[b_{ik}b_{kj}]_+ & \text{otherwise},
\end{cases}
\eeq
and ${\bf x}'=(x_j')$ with 
\beq \label{clustmut}
x_j'=
 \begin{cases}
x_k^{-1}\, f_k(M^+_k, M^-_k) &\text{for}  \,\,j=k  \\
 \ x_j   &\text{for}  \,\, j \neq k.
\end{cases}
\eeq 
Here, $[a]_+ =\max (a,0)$, 
$f_k:\mathbb{F} \times \mathbb{F}\rightarrow \mathbb{F}$ is a differentiable function and 
$$M^+_k:= 
\prod_{i=1}^N x_i^ {[b_{ki}]_+}
\;, \ M^-_k: =
\prod_{i=1}^N x_i^ {[-b_{ki}]_+}
\;.$$ 
For $f_k(M^+_k,M^-_k)=M^+_k+M^-_k$, the first  relation in  \eqref{clustmut} becomes the usual exchange relation 
$x_k'x_k=M^+_k+M^-_k$ for cluster mutations 
in a coefficient-free cluster algebra. In this case, we know that 
there is a log-canonical presymplectic form compatible with cluster mutations \cite{fg, gsv, in}. We extend this result to include more general types of  mutations. 


\begin{lem} \label{lem}
Let $Q$ be a quiver associated with the exchange matrix  $B=(b_{ij})$ and $(B', {\bf x}')={\mu}_k (B, {\bf x})$, as defined by \eqref{matmut} and \eqref{clustmut}. Then  
\begin{equation} \label{pres}
\sum_{i<j} \frac{b'_{ij}}{x'_i x'_j} \rd x'_i \wedge \rd x'_j=\sum_{i<j} \frac{b_{ij}}{x_i x_j} \rd x_i \wedge \rd x_j
\end{equation}
if and only if
\begin{equation} \label{functmut}
f_k(M^+_k,M^-_k)=M^+_k g_k\left(\frac{M^-_k }{M^+_k }\right)\;, 
\end{equation}
for an arbitrary differentiable function $g_k:\mathbb{F}\rightarrow \mathbb{F}$. 
\end{lem}

\begin{remark}\label{freedom} 
Equivalently, the function $f_k$ can be written in the form 
$$
f_k(M^+_k,M^-_k)=M^-_k \tilde{g}_k\left(\frac{M^+_k }{M^-_k }\right)\;, 
$$
for $\tilde{g}_k$ arbitrary.  
\end{remark}

\begin{prf}
Using  $\sum'$ to denote a sum over indices with index $k$ omitted, we 
have 
\begin{align*} 
\om & = \sum_{i<j} \frac{b_{ij}}{x_i x_j} \, \rd x_i \wedge \rd x_j 
\\
& = \tfrac{1}{2}\left(\text{\Large $\Sigma$}_{i,j}' b_{ij}\rd\log x_i\wedge \rd\log x_j 
+\text{\Large $\Sigma$}_i' b_{ik}
\rd\log x_i\wedge\rd\log x_k +\text{\Large $\Sigma$}_j' b_{kj}\rd\log x_k\wedge\rd\log x_j \right) \\
& =  \tfrac{1}{2} \text{\Large $\Sigma$}_{i,j}' b_{ij}\rd\log x_i\wedge \rd\log x_j   +  \text{\Large $\Sigma$}_i' b_{ik}
\rd\log x_i\wedge\rd\log x_k, 
\end{align*}
and similarly 
\begin{align*} 
\om ' & =  \sum_{i<j} \frac{b_{ij}'}{x_i' x_j'} \, \rd x_i' \wedge \rd x_j' 
\\
& =  \tfrac{1}{2} \text{\Large $\Sigma$}_{i,j}' b_{ij}'\rd\log x_i'\wedge \rd\log x_j'   +  \text{\Large $\Sigma$}_i' b_{ik}'
\rd\log x_i'\wedge\rd\log x_k' \\ 
& = 
 \tfrac{1}{2} \text{\Large $\Sigma$}_{i,j}' (b_{ij} +\text{sgn}(b_{ik})[b_{ik}b_{kj}]_+)\rd\log x_i\wedge \rd\log x_j  
 - \text{\Large $\Sigma$}_i' b_{ik}
\rd\log x_i\wedge(-\rd\log x_k +\rd\log f_k).   
\end{align*} 
Hence if we consider the sets 
$$\beta_k^+=\{i\in \{1,\dots N \}: b_{ki}>0 \}, \qquad  \beta_k^-=\{i\in \{1,\dots N \}: b_{ki}<0 \},$$ 
then noting that $[b_{ik}b_{kj}]_+=0$ unless either $i\in\beta_k^+$, $j\in\beta_k^-$ or vice versa, and 
defining 
$$ 
\rd S^\pm_k:=\pm \rd\log M_k^\pm=\sum_{i\in\beta_k^\pm}b_{ki}\rd\log x_i, 
$$ 
we have 
\begin{align*}
\om'-\om  &=   \tfrac{1}{2} \text{\Large $\Sigma$}_{i,j}' \text{sgn}(b_{ik})[b_{ik}b_{kj}]_+\rd\log x_i\wedge \rd\log x_j 
-  \text{\Large $\Sigma$}_i' b_{ik}
\rd\log x_i\wedge \rd\log f_k \\
 &=  \tfrac{1}{2}\left(\sum_{\substack{i\in \beta_k^- \\  j\in \beta_k^+}}b_{ik}b_{kj}\rd\log x_i\wedge \rd\log x_j 
- \sum_{\substack{i\in \beta_k^+ \\  j\in \beta_k^-}} b_{ik}b_{kj}\rd\log x_i\wedge \rd\log x_j \right) \\
&\quad +\text{\Large $\Sigma$}_i'b_{ki} \rd\log x_i\wedge 
\left(\frac{M^+_k}{f_k}\frac{\partial f_k}{\partial M^+_k}\rd\log M_k^+ + 
\frac{M^-_k}{f_k}\frac{\partial f_k}{\partial M^-_k}\rd\log M_k^-
\right) \\
& = -\sum_{\substack{i\in \beta_k^- \\  j\in \beta_k^+}}b_{ki}b_{kj}\rd\log x_i\wedge \rd\log x_j  \\ 
& \quad+  (\rd S^+_k +\rd S^-_k)\wedge \left(\frac{M^+_k}{f_k}\frac{\partial f_k}{\partial M^+_k}\rd S_k^+ -  
\frac{M^-_k}{f_k}\frac{\partial f_k}{\partial M^-_k}\rd  S_k^-
\right) \\ 
& = \left( \frac{M^+_k}{f_k}\frac{\partial f_k}{\partial M^+_k} + \frac{M^-_k}{f_k}\frac{\partial f_k}{\partial M^-_k} -1\right)\, \rd S_k^-\wedge\rd S_k^+.  
\end{align*} 
Hence $\om'=\om$ iff $f_k=f_k(M_k^+,M_k^-)$ satisfies the linear partial differential equation 
$$
M_k^+ \frac{\partial f_k}{\partial M_k^+ }+M_k^- \frac{\partial f_k}{\partial M_k^- }=f_k,
$$
of which the general solution is given by (\ref{functmut}) with $g_k$ arbitrary. 
\end{prf}

According to Lemma \ref{lem}, if the exchange matrix $B$ remains invariant under a sequence of mutations of the form \eqref{functmut} then the map that is generated 
by the same sequence of cluster mutations will preserve a presymplectic form, i.e.\ the following theorem holds.  

\begin{thm} \label{thm2}
Let ${\mu}_{i_1}, {\mu}_{i_2}, \dots,{\mu}_{i_\ell}$, for $i_j\in\{1,\dots,N\}$, $j\in \mathbb{N}$, be a sequence of mutations defined from 
\eqref{matmut} and \eqref{clustmut}, with each function $f_{i_j}$ being of the form   \eqref{functmut}, such that 
 $${\mu}_{i_\ell}\cdots{\mu}_{i_2} {\mu}_{i_1}(B,\mathbf{x})=(B,\tilde{\mathbf{x}}).$$
Then the map $\varphi:\mathbf{x}\mapsto\tilde{\mathbf{x}}$  
preserves the two-form
\begin{equation} \label{omega}
\omega=\sum_{i<j}^N \frac{b_{ij}}{x_i x_j} \rd x_i \wedge \rd x_j.
\end{equation}
\end{thm}

\begin{remark}\label{permperiodic} The preceding result admits a slight generalization to the case of cluster algebras (or quivers $Q$) with periodicity 
under mutations. In the most general setting, as described by Nakanishi \cite{nak}, these are defined by an exchange matrix with the property that ${\mu}_{i_\ell}\cdots{\mu}_{i_2} {\mu}_{i_1}(B)=\hat{\rho}(B)$, where 
$\hat{\rho}$ is some permutation of $(1,2,\ldots,N)$ acting on the indices (equivalently, on the nodes of the quiver $Q$). The particular case $\mu_m\cdots\mu_2\mu_1 (B)=\rho^m(B)$, for the cyclic permutation 
$\rho: (1,2,\ldots,N)\mapsto (N,1,2,\ldots, N-1)$ was called cluster mutation-periodicity with period $m$ by Fordy and Marsh \cite{FM}, who gave a complete classification of the case $m=1$. 
A straightforward adaptation of the above argument shows that if $B$ is periodic, then the map $\varphi=\hat{\rho}^{-1} {\mu}_{i_\ell}\cdots{\mu}_{i_2} {\mu}_{i_1}$ leaves $B$ invariant 
and preserves the corresponding log-canonical presymplectic form (\ref{omega}), in the sense that $\varphi^*(\om)=\om$. Lemma 2.3 in \cite{FH} covers the special case of 
this result for ordinary cluster mutations when $B$ is cluster mutation-periodic with period 1, so   
$\varphi=\rho^{-1}\mu_1$ and the map can be written as a single recurrence relation. We shall consider an example of this with a generalized mutation in section \ref{sgredn}. 
The slightly different (but closely related) problem of when an ordinary difference equation 
preserves a   log-canonical Poisson bracket was considered in \cite{eqr}.  
\end{remark}

In the next section our aim is to generalize the example of the Lyness map (\ref{ly}), corresponding to the root system $A_2$,  to other finite type root systems of type $A$, by taking 
mutations defined by affine functions $f_k$ with additional parameters that destroy the Laurent property but preserve the two-form (\ref{omega}). Section 3 contains  more general choices of mutations, starting from 
affine Dynkin diagrams, where 
the factors $g_k$ in  \eqref{functmut} involve M\"obius transformations, which lead to travelling wave reductions of the discrete sine-Gordon equation. We end with a few final remarks.

\section{Deformations of type $A$ periodic maps} 
\setcounter{equation}{0}

In this section, extra parameters are included in the regular exchange relation by taking 
$g_k(x)=b_k x+a_k$, since 
\beq\label{fk}
f_k(M^+_k,M^-_k)=M^+_k g_k\left(\frac{M^-_k }{M^+_k }\right)=a_k M^+_k+b_kM^-_k\;. 
\eeq
Hence,  according to Theorem \ref{thm2}, quivers which are periodic under a particular sequence of mutations 
(or more generally, are periodic up to a permutation) 
give rise to 
 parametric cluster maps that preserve the presymplectic form  \eqref{omega}. If the corresponding exchange matrix is non-singular 
 the parametric cluster maps are symplectic. We begin by examining the case of $A_2$ in more detail, 
and then apply this approach to study the integrability of parametric cluster maps associated with the $A_3$ and $A_4$ quivers.  
  
\subsection{Deformed mutations for $A_2$ quiver} 

The exchange matrix of type $A_2$ is  
\begin{equation*}
B= \left( \begin{array}{rr} 0 & 1  \\
-1 & 0 
\end{array} \right). 
\end{equation*}
In this case,  $B$ corresponds to a cluster mutation-periodic quiver with period $1$ and $M^+_1=x_2$, $M^-_1=1$. 
So, by the modification of Theorem \ref{thm2} as in Remark \ref{permperiodic},  taking $\rho: (1,2)\mapsto (2,1)$,   for any differentiable function $\tilde{g}:\mathbb{F} \rightarrow \mathbb{F}$ 
the map $\varphi=\rho^{-1}\mu_1$ given by 
\begin{equation} \label{A2map}
\varphi:(x_1,x_2)\mapsto \left(x_2,\frac{1}{x_1}\tilde{g}(x_2)\right)\;,
\end{equation}
is symplectic with respect to $\omega=\frac{1}{x_1 x_2} \rd x_1 \wedge \rd x_2$. (Compared with (\ref{functmut}) we have 
$f_1(x,1)=xg_1(1/x)=\tilde{g}(x)$: in general, replacing $g_k(x)\to xg_k(1/x)$ corresponds to sending $B\to -B$, which 
is equivalent to replacing the corresponding quiver $Q\to Q^{opp}$, the same quiver with all arrows reversed; see also Remark \ref{freedom}.) 

With $(x,y)=(x_1,x_2)$ and $\tilde{g}(x)=ax+b$, we reproduce the Lyness map (\ref{ly}). Starting from the periodic map (\ref{ly5}), and relabelling the 
initial data as $(x_0,x_1)$, any cyclic symmetric function of 
the iterates $x_0,x_1,x_2,x_3,x_4$ in the periodic orbit (\ref{5cycle}) gives a first integral. So in the periodic case there are two independent integrals, namely 
\begin{align*}
K_1= &\sum_{j=0}^4 x_j=-3+\prod_{j=0}^4 x_j =  \frac{x_0^2x_1+x_0x_1^2 +x_0^2+x_1^2+2(x_0+x_1)+1}{x_0x_1}, \\
K_2=&\,\sum_{j=0}^4 x_j x_{j+1} \\
=& \,
\frac{x_0x_1(x_0^2x_1^2+x_0^3+x_1^3+x_0^2+x_1^2+x_0+x_1+2) +x_0^3+x_1^3 
+2(x_0^2+x_1^2)+x_0+x_1} 
{x_0^2x_1^2}. 
\end{align*}
Both of the latter are sums of Laurent monomials, so in the case of the map with parameters, first integrals 
can be sought by taking arbitrary linear combinations of the same monomials and solving the resulting conditions 
on the coefficients. Thus in the case of (\ref{ly}), the first integral (\ref{integral}) can be considered as a deformation 
of $K_1$ above; but a first integral composed of the Laurent monomials in $K_2$ only exists when $b=a^2$ and 
the map is periodic, corresponding to the undeformed situation. 

Although the Laurent phenomenon does not persist for the iterates of the Lyness recurrence 
\beq\label{lrec}
x_{n+2}x_n=ax_{n+1}+b
\eeq 
 when $b\neq a^2$, it was pointed out in \cite{FH} 
that there is a connection to a   cluster algebra via a lift to a space of higher dimension, 
defined by the substitution 
$$ 
x_n = \frac{\tau_{n+5}\tau_n}{\tau_{n+3}\tau_{n+2}}, 
$$
which leads to the Somos-7 recurrence 
\beq\label{s7}
\tau_{n+7}\tau_n =a\, \tau_{n+6}\tau_{n+1}+b\,\tau_{n+4}\tau_{n+3}.
\eeq 
As explained in \cite{FM}, Somos-type recurrences such as the above, with a sum of two monomials on the 
right-hand side, can be generated by mutations in a cluster algebra. In the case of (\ref{s7}), it is a cluster 
algebra of rank 7, extended by the addition of the parameters $a,b$ as frozen variables. 

The rest of this section is devoted to the analogous constructions for $A_3$ and $A_4$.

\subsection{$A_3$ quiver with parameters}
For the $A_3$ quiver with exchange matrix 
\begin{equation*}
B= \left( \begin{array}{ccc} 0 & 1 & 0  \\
-1 & 0 & 1 \\
0 & -1 & 0  \\
\end{array} \right),
\end{equation*}
as in Figure 1, 
we take $f_k(M^+_k,M^-_k)=a_k M^+_k+b_kM^-_k$.
In this case, 
$$\varphi(B,{\bf x}):={\mu}_{3}{\mu}_{2}{\mu}_{1}(B,{\bf x})=\big(B,\varphi({\bf x})\big),$$
where the composition $\varphi=\mu_3\mu_2\mu_1$ acts on the cluster variables 
${\bf x}=(x_1,x_2,x_3)$ according to 
\beq \label{A3maps}\begin{array}{rcl}
\mu_1: \quad (x_1,x_2,x_3)\mapsto (x_1',x_2,x_3), \qquad x_1'x_1 & = & b_1+a_1 x_2, \\
 \mu_2: \quad (x_1',x_2,x_3)\mapsto (x_1',x_2',x_3), \qquad x_2'x_2 & = & b_2+a_2 x_1'x_3, \\ 
\mu_3: \quad (x_1',x_2',x_3)\mapsto (x_1',x_2',x_3'), \qquad x_3'x_3 & = & b_3+a_3 x_2'. 
\end{array}
\eeq 
Since 
$\varphi(B)=B$, so the exchange matrix $B$ remains invariant under this sequence of mutations,   
by Theorem \ref{thm2} the map $\varphi$ preserves the corresponding log-canonical two-form, that is 
$$ 
\varphi^*(\om)=\om, 
$$  
where 
\begin{equation*} \label{omegaA4}
\omega=\frac{1}{x_1 x_2} \rd x_1 \wedge \rd x_2+\frac{1}{x_2 x_3} \rd x_2 \wedge \rd x_3\;.
\end{equation*}

\begin{figure}
 \centering
\label{quiver0}
\epsfig{file=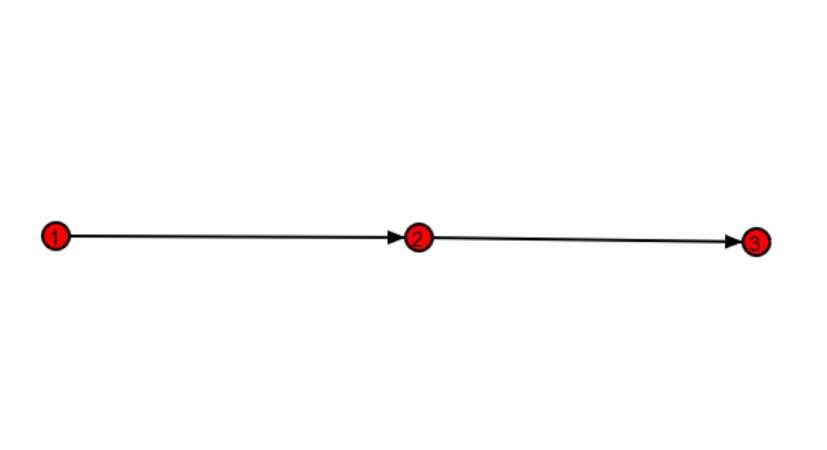, height=1.2in, width=2.5in}
\caption{The $A_3$ quiver.} 
\end{figure} 

The original coefficient-free cluster algebra is given by setting $a_i=1=b_i$ for $i=1,2,3$, 
and  in that case the map $\varphi$ is periodic with period 6, that is $\varphi^6({\bf x}) ={\bf x}$.  Moreover, one can write 
down three independent first integrals for the periodic map, by taking appropriate  symmetric functions along each orbit, such as 
$\sum_{i=0}^5 (\varphi^*)^i(x_j)$, $\prod_{i=0}^5 (\varphi^*)^i(x_j)$, etc.

However, before considering the deformed case (\ref{A3maps}), there are two ways to simplify the calculations. First of 
all, assuming the case of generic parameter values $a_ib_i\neq 0$ for all $i$, we apply the scaling action of the three-dimensional algebraic torus $(\F^*)^3$, given by $x_i\to \la_i\, x_i$, $\la_i\neq0$, and use this to remove three parameters, so that we 
obtain 
$$ 
a_1\to 1,\quad b_1\to 1, \quad a_2\to d, \quad b_2\to c,\quad a_3\to 1,\quad b_3\to e, 
$$ 
where $c,d,e$ are arbitrary. Having simplified the space of parameters, the map $\varphi$ is equivalent to iteration of the 
system of recurrences 
\beq\label{A3recs} 
\begin{array}{rcl}
x_{1,n+1}x_{1,n} & = & x_{2,n}+1, \\  
x_{2,n+1}x_{2,n} & = & dx_{1,n+1}x_{3,n}+c, \\  
x_{3,n+1}x_{3,n} & = & x_{2,n+1}+e.   
\end{array} 
\eeq 
Secondly, because we are in an odd-dimensional situation where $B$ necessarily has determinant 
zero, so that $\om$ is degenerate, so following \cite{FH} (cf. Theorem 2.6 therein) we can use 
$$ 
\mathrm{ker}\,B=<(1,0,1)^T>, \qquad 
\mathrm{im}\, B=  (\mathrm{ker}\,B)^\perp = <(0,1,0)^T, (-1,0,1)^T>
$$ 
to generate the one-parameter scaling group $(x_1,x_2,x_3)\to (\la x_1,x_2,\la x_3)$ and the projection $\pi$ onto its monomial invariants,  
$$ 
\pi: \qquad 
y=x_2, \qquad w=\frac{x_3}{x_1}.  
$$ 
On the $y,w$-plane, $\varphi$ induces the 
reduced map 
\beq\label{phihat} 
\hat{\varphi}: \qquad
\left(\begin{array}{c} y\\ w \end{array}\right) \mapsto 
\left(\begin{array}{c} \big(d(y+1)w+c\big)/y\\ (dw+c)/(y w) +(e-c)/\big(w(y+1)\big)\end{array}\right),
\eeq 
which is symplectic, preserving the nondegenerate two-form 
\beq\label{omhat} 
\hat{\om}=\rd \log y \wedge \rd \log w, \qquad \pi^*\hat{\om}=\om. 
\eeq 

In the original case where all parameters are 1, the reduced map (\ref{phihat}) with $c=d=e=1$ has period 3, because 
$x_{2,n+3}=x_{2,n}$ and $x_{3,n+3}/x_{1,n+3}=x_{3,n}/x_{1,n}$ for all $n$. Thus  in that case there are two 
functionally independent first integrals in the plane, which can be taken as 
\beq\label{A3ints} \begin{array}{rcccl}
K_1&=&\prod_{i=0}^2 (\hat{\varphi}^*)^i(y)&=&\frac{(yw+w+1)(y+w+1)}{yw}
=-2 +\sum_{i=0}^2 (\hat{\varphi}^*)^i(y), \\
K_2&=&\sum_{i=0}^2 (\hat{\varphi}^*)^i(w)&=&\frac{yw^3+yw^2+y^2w+w^2+2w+1}{yw(w+1)}
\end{array} 
\eeq  
(while the product $\prod_{i=0}^2 (\hat{\varphi}^*)^i(w)=1$, so does not give a nontrivial integral).

Next, we modify $K_1$ and $K_2$ by inserting constant coefficients in front of each of their terms, which are all Laurent monomials in $K_1$, while for $K_2$ we can replace the term $w+1$ in the denominator by an arbitrary linear function 
of $w$. If we require that (at least) one of these modified integrals should be preserved by the map $\hat{\varphi}$, then 
this puts a finite number of constraints on the coefficients and parameters $c,d,e$, which are necessary and sufficient for the deformed symplectic map 
to be Liouville integrable.  
Thus we obtain the following result.

\begin{thm} 
The condition 
$$c=e$$ is necessary and sufficient for the symplectic map (\ref{phihat}) to admit a deformation of 
the first integral $K_1$, given by 
\beq\label{newk1}
 K_1=\frac{(yw+w+d)(y+dw+c)}{yw}, 
\eeq 
hence $\hat{\varphi}$ is integrable whenever this condition holds. 
Requiring that a deformation of $K_2$ should be preserved imposes the stronger conditions 
$$
c=d^2=e, 
$$ 
in which case both 
\beq\label{newk2}
K_2=\frac{w^3y+d(y+1)w^2+(y^2+2d^2)w+d^3}{yw(w+d)} 
\eeq 
and $K_1$ given by (\ref{newk1}) with $c=d^2$ are preserved, and all the orbits of $\hat{\varphi}$ are 
periodic with period 3.
\end{thm}
\begin{prf} 
Starting from a general sum of monomials 
$$ 
K_1= y +\al\, w +\be\, \frac{w}{y}  +\frac{\gam}{y}+\frac{\delta}{w}+\frac{\eps}{yw}+\mathrm{const}
$$  
(where we have fixed the scale by assuming that the first term has coefficient 1, and there is 
the freedom to add an arbitrary constant), 
we apply the map (\ref{phihat}) and require that $\hat{\varphi}^*(K_1)=K_1$. Comparing the rational functions 
one each side of the latter equation imposes the requirement $c=e$ and fixes $\al=\be=d$, $\gam=c+d^2$, 
$\delta=d$, $\eps=cd$; then choosing to add the constant $c+1$ means that $K_1$ can be factored as in (\ref{newk1}). 
Applying the same approach to $K_2$ requires the additional constraint $c=d^2$, restricting to the  
one-parameter family of period 3 maps 
$$ 
\hat{\varphi}: \qquad
\left(\begin{array}{c} y\\ w \end{array}\right) \mapsto 
\left(\begin{array}{c} \big(d(y+1)w+d^2\big)/y\\ d(w+d)/(y w) \big)\end{array}\right),
$$ 
which have two independent first integrals
given by (\ref{newk1}) with $c=d^2$ and (\ref{newk2}).
\end{prf} 

\begin{remark}
When $c=e$, the integrable symplectic map 
 \beq\label{phihati} 
\hat{\varphi}: \qquad
\left(\begin{array}{c} y\\ w \end{array}\right) \mapsto 
\left(\begin{array}{c} \big(d(y+1)w+c\big)/y\\ (dw+c)/(y w)\end{array}\right),
\eeq 
preserves the pencil of biquadratic curves defined by (\ref{newk1}), which means that 
there is a map of QRT type \cite{duistermaat, qrt} preserving the same pencil, given by the composition of the horizontal and vertical 
switch on each curve in the pencil, namely 
\beq\label{qrtpen} 
\hat{\psi}: \qquad 
\left(\begin{array}{c} y\\ w \end{array}\right) \mapsto  
\left(\begin{array}{c} \bar{y}\\ \bar{w} \end{array}\right), \qquad \bar{y}y=\frac{(dw+c)(w+d)}{w}, 
\quad \bar{w}w=\frac{\bar{y}+c}{\bar{y}+1}.   
\eeq 
From general considerations about automorphisms of elliptic curves, since they each correspond to 
translation by a point, these two maps should commute with one another, 
and indeed it is straightforward to verify that 
$$ 
\hat{\psi}\circ \hat{\varphi}=\hat{\varphi}\circ\hat{\psi}.
$$ 
However, it appears that generically the two maps correspond to translation by two independent points 
of infinite order, so (over $\Q$, say) this should generate a family of curves with Mordell-Weil group of rank at least 2.  
(As a special case, when $c=d=1$ the map $\hat{\psi}$ has period 2 for any initial data, corresponding to translation by a 
2-torsion point, whereas the period 3 map $\hat{\varphi}$ corresponds to addition of a 3-torsion point; so the points are independent, albeit 
not of infinite order in this case.)
\end{remark}

We now treat the singularity pattern of the iterates of (\ref{phihati}), in order to obtain its Laurentification 
in the sense of \cite{hhkq}, i.e.\ a lift to a map with the Laurent property in a space of higher dimension, in which the new variables can be regarded as tau functions.   Rather than a standard singularity confinement analysis, we 
study orbits defined over $\Q$, and   consider a $p$-adic analogue of confinement, as in \cite{kanki}. The possible singularity patterns can then be obtained using the empirical approach introduced in \cite{hkq}, simply by inspecting 
the prime factorization of a few terms along a particular orbit. 

Thus we choose some particular values for the 
coefficients and initial data: taking  $c=2$, $d=3$ and $(y_0,w_0)=(1,1)$, we find the first   
few iterates are 
$$ 
(8,5),(\tfrac{137}{8},\tfrac{17}{40}),(\tfrac{1607}{1096},\tfrac{1048}{2329}), 
(\tfrac{800200}{220159},\tfrac{1068874}{210517}),(\tfrac{3210496223}{160740175},\tfrac{728705399}{780395050}),
(\tfrac{7129742296469}{2344013756975}, \tfrac{2735651842025}{10626437852503}),
$$ 
so that the  values of $y_n$ for $n=1,2,3,\ldots$ factorize as 
$$ 
2^3, \tfrac{137}{2^3},\tfrac{1607}{2^3\cdot 137}, \tfrac{2^3\cdot 5^2 \cdot 4001}{137\cdot 1607}, 
\tfrac{11\cdot 17\cdot 113\cdot 137\cdot 1109}{5^2\cdot 1607\cdot 4001}, 
\tfrac{13\cdot 19\cdot 43\cdot 1607\cdot 417727}{5^2\cdot 11\cdot 17\cdot 113\cdot 1109\cdot 4001},
\ldots ,
$$ while the factorizations of the corresponding values of $w_n$ are 
$$ 
5, \tfrac{17}{2^3\cdot 5},\tfrac{2^3\cdot 131}{17\cdot 137}, \tfrac{2\cdot 47\cdot 83\cdot 137}
{131\cdot 1607}, \tfrac{467\cdot 971\cdot 1607}{2\cdot 5^2\cdot 47\cdot 83\cdot 4001}, 
\tfrac{5^2\cdot 4001\cdot 27349681}{11\cdot 17\cdot 113\cdot 467\cdot 971\cdot 1109}, 
\ldots, 
$$ 
and so on. For the primes $p=113, 137,1607,4001$, the values of the $p$-adic norm $|y_n|_p$ follow the 
pattern $1,p^{-1},p,p,p^{-1},1$, with the corresponding values of $|w_n|_p$ being $1,1,p,p^{-1},1,1$, 
while for the primes $p=2$ and $5$ there are instances of the same patterns but with $p\to p^3$ and $p\to p^2$, 
respectively. (For some of these primes, the whole pattern is not visible above, but  it can easily be verified by computing 
the next few terms, which are omitted here.) In $w_n$ there are also other primes that do not appear in $y_n$, 
e.g.\ $p=17,47,83,131,467,971$, and for these the pattern of $|w_n|_p$ is $1,p^{-1},p,1$. This immediately 
suggests that $y_n,w_n$ can be written using two different tau functions $\si_n,\tau_n$, as 
\beq\label{tauf} 
\tilde{\pi}: \qquad 
y_n=\frac{\tau_{n-2}\tau_{n+1}}{\tau_{n-1}\tau_n}, \quad 
w_n = \frac{\si_{n+1}\tau_{n-1}}{\si_n\tau_{n}}, 
\eeq 
so that the first type of $p$-adic singularity corresponds to $\tau_n\equiv 0\bmod  p$ for some $n$, and the second occurs 
when $\si_n\equiv 0\bmod  p$.

Our next goal is to show that the tau functions in (\ref{tauf}) satisfy a system of bilinear equations, namely 
\beq\label{tausys} 
\begin{array}{rcl}
\si_{n+2}\tau_{n-2}& =& d\,\si_{n+1}\tau_{n-1}+c\, \si_n\tau_n, \\ 
\si_n \tau_{n+2} & = & \si_{n+2}\tau_n+ d\, \si_{n+1}\tau_{n+1} 
\end{array} 
\eeq 
(we expect that these could be 
viewed as a reduction of coupled discrete Hirota equations \cite{dl, zabrodin}), 
and to prove that this system has the Laurent property. The first equation 
in (\ref{tausys}) is straightforward to obtain, as it arises directly from substituting the tau function 
expressions (\ref{tauf})  into the second component of (\ref{phihati}), rewritten in the form of a recurrence, 
but the second bilinear equation requires more work. If we look at the singularity pattern 
in the original three-dimensional system (\ref{A3recs}) with $e=c$, then we see that 
$$ 
x_{1,n}=\rho_n \, \frac{\si_{n+1}}{\tau_n}, \qquad 
x_{3,n}=\rho_n \, \frac{\si_{n}}{\tau_{n-1}},
$$ 
with a new prefactor $\rho_n$ appearing, while $x_{2,n}=y_n$ is already accounted for. Substituting in these 
formulae to rewrite the system (\ref{A3recs}) in terms of $\rho_n,\si_n,\tau_n$ yields 
\beq\label{tril} \begin{array}{rcl}
\rho_n\rho_{n+1}\, \si_{n+1}\si_n& =& \tau_{n+1}\tau_{n-2}+\tau_n\tau_{n-1}, \\ 
\tau_{n+2}\tau_{n-2}& =& \rho_n\rho_{n+1}\, d\, \si_{n+1}^2 + c\,\tau_n^2, \\ 
\rho_n\rho_{n+1}\, \si_{n+2}\si_{n+1}& =&  \tau_{n+2}\tau_{n-1}+c\,\tau_{n+1}\tau_{n}.
\end{array}  
\eeq 
For the above system, the initial values are $\rho_0,\si_0,\si_1, \tau_{-2},\tau_{-1},\tau_0,\tau_1$, and in principle 
one could use this to give a direct proof
that the sequences $(\si_n)$ and $(\tau_n)$ are Laurent polynomials in the 
initial data, although the sequence $\rho_n$ is not. However, note that the product  $\rho_n\rho_{n+1}$ can 
be eliminated from any two of the equations in (\ref{tril}), so doing this for each pair gives a set of three equations 
of degree 3, and then eliminating $\tau_{n+2}$ from any two of the latter results in the first equation in (\ref{tausys}), while  eliminating $\tau_{n+2}$ instead produces the relation 
$$ 
\si_n\tau_{n+2}\tau_{n-2}=d\,\si_{n+1}(\tau_{n+1}\tau_{n-2}+\tau_n\tau_{n-1})+c\,\si_n\tau_n^2.
$$ 
Finally, the second relation in (\ref{tausys}) follows by combining the first relation with the above to eliminate 
$\tau_{n-2}$. 

Immediate evidence for the Laurent property can be seen by iterating the system (\ref{tausys}) for $c=2$, $d=3$ with all 
initial values $\tau_{-2}=\tau_{-1}=\tau_0 = \tau_1=\si_0=\si_1=1$, corresponding to the initial values $y_0=w_0=1$ in the orbit 
considered above. The first few terms are the integers 
\small
$$ 
\begin{array}{rllllllll}
(\tau_n)_{n\geq 1}: 
\, &
1, & 8, &  
137, &
1607, &
100025, &
23434279, &
4436678467, &
1750170148834,   
\\
(\si_n)_{n\geq 1}: &
1, & 5, & 17, & 131, & 7802, & 453457, & 27349681,  & 18332191183, 
\end{array} 
$$
\normalsize  
and so on.
It is also easy to verify directly that the first few terms $\tau_2,\si_1$, etc.\ obtained by iteration of (\ref{tausys}) are Laurent polynomials in the initial data with coefficients 
belonging to $\Z[c,d]$.  

\begin{figure}
 \centering
\label{quiver1}
\epsfig{file=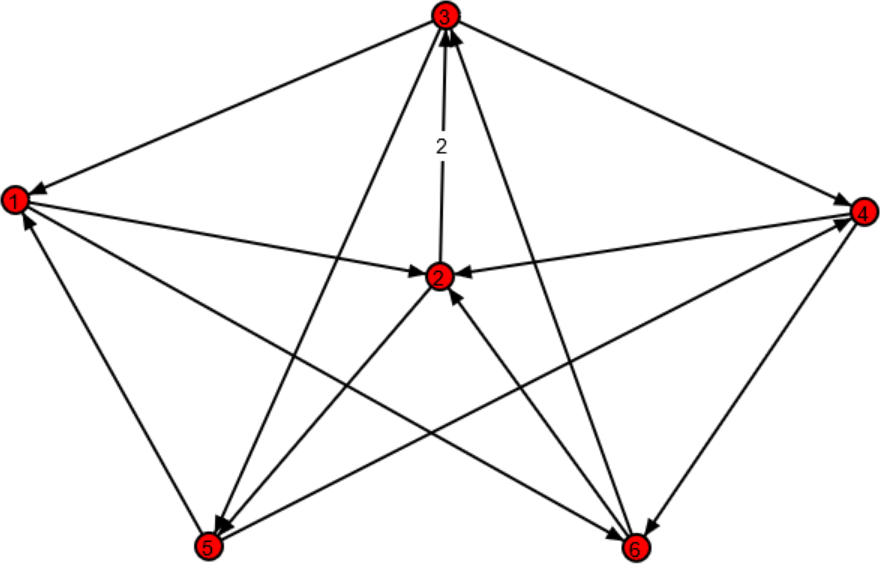, height=3in, width=3in}
\caption{The initial quiver $Q$ associated with the exchange matrix (\ref{BtauA3}).} 
\end{figure} 

To make further progress, it is helpful to consider the initial data for   (\ref{tausys}) as a set 
of cluster variables $(\tx_1,\tx_2,\tx_3,\tx_4,\tx_5,\tx_6)=(\tau_{-2},\tau_{-1},\tau_0,\tau_1,\si_0,\si_1)$, and calculate the pullback of the symplectic form (\ref{omhat}) by the map $\tilde{\pi}$ defined by the 
tau function expressions (\ref{tauf}), that is 
\beq\label{bstar}
\tilde{\om}=\tilde{\pi}^*\hat{\om}=\sum_{i<j} b_{ij}^*\rd \log \tx_i \wedge \rd \log \tx_j, 
\eeq  
where $B^*=(b_{ij}^*)$ is the skew-symmetric matrix
\beq\label{BtauA3}
B^*=\left(\begin{array}{cccccc}
0 & 1& -1 & 0 & -1 & 1 \\ 
-1 & 0 & 2 & -1 & 1 & -1 \\ 
1 & -2 & 0 & 1 & 1 & -1 \\ 
0 & 1 & -1 & 0 & -1 & 1 \\ 
1 & -1 & -1 & 1 & 0 & 0 \\ 
-1 & 1 & 1 & -1 & 0 & 0 
\end{array}
\right).
\eeq 
The quiver corresponding to this matrix is shown in Figure 2. 
 It is not hard to see that, when $c=1=d$, the bilinear equations (\ref{tausys}) for $n=0$ 
are generated by applying  a mutation at node 1, denoted by $\tmu_1$ (to distinguish it from 
mutations in the original $A_3$ quiver), followed by mutation $\tmu_5$: see Figure 3. 
To prove the Laurent property for the  case of arbitrary coefficients, it is necessary to extend the quiver with 
two extra frozen nodes.

\begin{figure}
\centering
\begin{subfigure}{.5\textwidth}
  \centering
  \includegraphics[width=.7\linewidth]{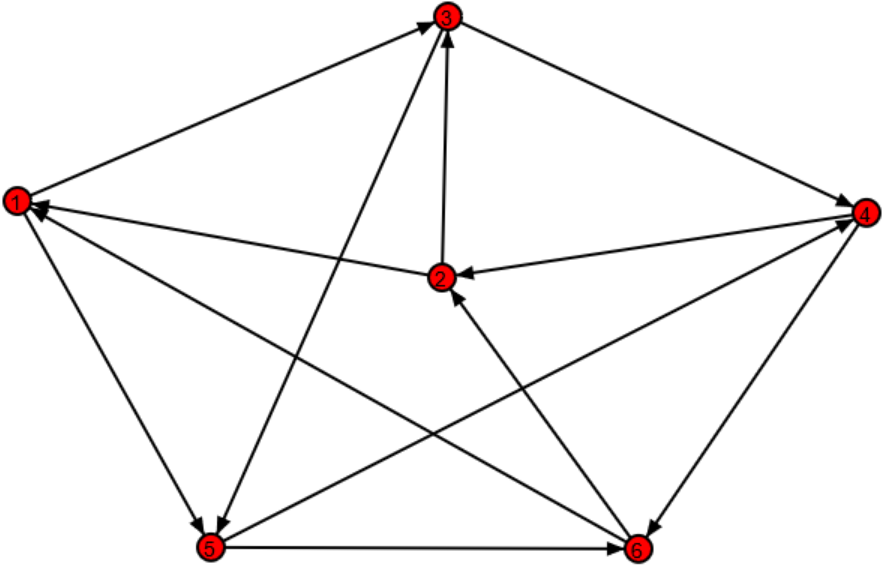}
  \caption{The quiver $\tilde{\mu}_1(Q)$.}
  \label{quiver2} 
\end{subfigure}%
\begin{subfigure}{.5\textwidth}
  \centering
  \includegraphics[width=.7\linewidth]{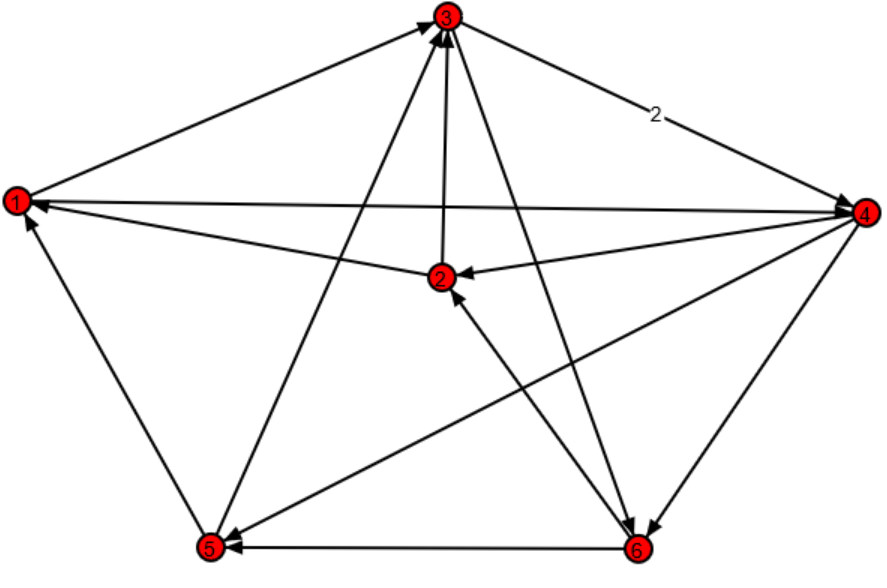}
  \caption{The quiver $\tmu_5\tmu_1(Q)$.}
  \label{quiver3} 
\end{subfigure}
\caption{The effect of two mutations on the quiver corresponding to (\ref{BtauA3}).}
\label{A3taumut}
\end{figure}

\begin{thm}\label{LPA3taus}
The sequences of tau functions $(\si_n)$ and $(\tau_n)$ for the integrable map 
(\ref{phihati}) consist of elements of the Laurent polynomial ring 
$\Z_{>0}[c,d,\tau_{-2}^{\pm 1},\tau_{-1}^{\pm 1},\tau_{0}^{\pm 1},\tau_{1}^{\pm 1},\si_{0}^{\pm 1},\si_{1}^{\pm 1}]$,  being generated 
by a sequence of mutations in a cluster algebra defined by the quiver in Figure 2 with the addition of two frozen nodes. 
\end{thm}

\begin{prf}
In order to include the coefficients, we define an 
extended cluster  $\tilde{\bf x}=(\tx_1,\ldots,\tx_8)=(\tau_{-2},\ldots,\tau_1,\si_0,\si_1,c,d)$,
where $\tx_7=c$ and $\tx_8=d$ are frozen 
variables, and take an extended exchange matrix   
\beq\label{BtauA3ext}
\tilde{B}^*=\left(\begin{array}{cccccc}
0 & 1& -1 & 0 & -1 & 1 \\ 
-1 & 0 & 2 & -1 & 1 & -1 \\ 
1 & -2 & 0 & 1 & 1 & -1 \\ 
0 & 1 & -1 & 0 & -1 & 1 \\ 
1 & -1 & -1 & 1 & 0 & 0 \\ 
-1 & 1 & 1 & -1 & 0 & 0 \\ 
1 & 0 & 0 & 0 & 0 & -1 \\ 
-1 & -1 & 1 & 1 & 0 & 0 
\end{array}
\right),
\eeq 
where  two more rows have been appended to (\ref{BtauA3}). 
(The diagram of the quiver with the additional arrows to/from the frozen nodes 
does not look quite so clear compared with Figure 2, so it has been omitted.) 
Applying the mutation $\tmu_1$ gives the exchange relation 
$$ 
\si_2\tau_{-2}=d\,\si_1\tau_{-1}+c\,\si_0\tau_0, 
$$ 
and produces a new cluster $(\si_2,\tau_{-1},\tau_0,\tau_1,\si_0,\si_1,c,d)$ 
and a new matrix $\tmu_1(\tilde{B}^*)$ corresponding to the quiver in Figure 3(a) 
with appropriate arrows to/from the frozen nodes 7 and 8. 
Next, by applying  the mutation $\tmu_5$, the exchange relation is 
$$ 
\tau_2\si_0=d\, \si_1\tau_1+\si_2\tau_0, 
$$ 
with the new cluster being 
 $(\si_2,\tau_{-1},\tau_0,\tau_1,\tau_2,\si_1,c,d)$, and the new exchange 
matrix $\tmu_5\tmu_1(\tilde{B}^*)$ corresponding to the quiver in Figure 3(b) 
with suitable extra arrows added to take the coefficients into account. 
Continuing in a similar way, we find a sequence of mutations to successively generate 
$\si_3,\tau_3,\si_4,\tau_4$, and so on, such that overall after applying the composition 
of 12 mutations given by 
\beq\label{munotn}
\tmu_{463524136251}:=\tmu_4\tmu_6\tmu_3\tmu_5\tmu_2\tmu_4\tmu_1\tmu_3\tmu_6\tmu_2\tmu_5\tmu_1
\eeq
(in order from right to left), the quiver returns to its starting position;  so we have 
$$ 
 \tmu_{463524136251}(\tilde{B}^*)= \tilde{B}^*, \qquad 
 \tmu_{463524136251}(\tilde{\bf x})=(\tau_4,\tau_5,\tau_6,\tau_7,\si_6,\si_7,c,d),
$$ 
with the index of each of the tau functions increased by 6. Hence by induction both sequences $(\si_n)$, $(\tau_n)$ are generated 
by repeatedly applying this composition of mutations, and the Laurent property follows from the fact that the tau functions are all 
elements of the cluster algebra, for which it is also known that the Laurent polynomials in the initial data have positive integer coefficients \cite{ghkk, ls}.   
\end{prf}
\begin{remark}
Preliminary calculations suggest that the iterates of the QRT map (\ref{qrtpen}), which commutes with $\hat{\varphi}$, have a different singularity structure, 
corresponding to a tau function substitution of the form 
$$ 
y_n=\frac{\eta_{n}}{\si_n\tau_{n-1}}, \quad 
w_n = \frac{\si_{n+1}\tau_{n-1}}{\si_n\tau_{n}}, 
$$ 
where $\eta_n$ has weight two. It would be interesting to see whether this has a cluster algebra interpretation.
\end{remark}

\subsection{$A_4$ quiver with parameters}
For the exchange matrix 
\begin{equation*}
B= \left( \begin{array}{cccc} 0 & 1 & 0 & 0 \\
-1 & 0 & 1 & 0 \\
0 & -1 & 0 & 1 \\
0 & 0 & -1 & 0
\end{array} \right),
\end{equation*}
corresponding to the quiver of type $A_4$, once again we start from 
functions of the form $f_k(M^+_k,M^-_k)=a_k M^+_k+b_kM^-_k$, with 
arbitrary coefficients such that $a_kb_k\neq 0$. By rescaling $x_j\to \la_j \, x_j$ with $\la_j\in\F^*$, we can set four of the parameters to 1, so that it is sufficient 
to consider a four-parameter family of mutations, given by 
\beq \label{A4maps}\begin{array}{rcl}
\mu_1: \quad (x_1,x_2,x_3,x_4)\mapsto (x_1',x_2,x_3,x_4), \qquad x_1'x_1 & = & b_1+a_1 x_2, \\
 \mu_2: \quad (x_1',x_2,x_3,x_4)\mapsto (x_1',x_2',x_3,x_4), \qquad x_2'x_2 & = & 1+ x_1'x_3, \\ 
\mu_3: \quad (x_1',x_2',x_3,x_4)\mapsto (x_1',x_2',x_3',x_4), \qquad x_3'x_3 & = & 1+ x_2'x_4, \\ 
\mu_4: \quad (x_1',x_2',x_3',x_4)\mapsto (x_1',x_2',x_3',x_4'), \qquad x_4'x_4 & = & b_4+ a_4 x_3'. 
\end{array}
\eeq 
Then,  defining the action of $\varphi=\mu_4\mu_3\mu_2\mu_1$ on the cluster ${\bf x}=(x_1,x_2,x_3,x_4)$ as above, 
$$\varphi(B,{\bf x}):={\mu}_{4}{\mu}_{3}{\mu}_{2} {\mu}_{1}(B,{\bf x}) =\big(B,\varphi({\bf x})\big),$$
so the nondegenerate exchange matrix $B$ remains invariant under this sequence of mutations,   and 
according to Theorem \ref{thm2} the map 
\begin{equation*}
{\bf x} \mapsto\varphi({\bf x})
\end{equation*}
is symplectic with respect to 
\begin{equation} \label{omegaA4}
\omega=\frac{1}{x_1 x_2} \rd x_1 \wedge \rd x_2+\frac{1}{x_2 x_3} \rd x_2 \wedge \rd x_3+\frac{1}{x_3 x_4} \rd x_3 \wedge \rd x_4\;.
\end{equation}
Equivalently, by computing the inverse matrix $P=B^{-1}=(p_{ij})$, the map $\varphi$ preserves the nondegenerate Poisson bracket given by 
$\{\,x_i,x_j\,\}=p_{ij}\,x_ix_j$, which has the explicit form 
\beq\label{bracket} 
\{\,x_2,x_1\,\}=x_2x_1, \qquad \{\,x_4,x_1\,\}=x_4x_1, \qquad \{\,x_4,x_3\,\}=x_4x_3, 
\eeq 
with all other brackets being zero.

In the original case of the undeformed quiver, corresponding to $a_1=a_4=b_1=b_4=1$ in (\ref{A4maps}), the map $\varphi$ is completely periodic with period 7, 
and admits four independent integrals in dimension four. Here we focus on 
\beq\label{A4ints}
I_1=\sum_{j=0}^6(\varphi^*)^j(x_1), \qquad  I_2=\prod_{j=0}^6(\varphi^*)^j(x_1),
\eeq 
since in the undeformed case these Poisson commute with respect to the bracket (\ref{bracket}), that is 
\beq\label{invol}
\{\, I_1,I_2\,\}=0. 
\eeq
Being a sum/product of cluster variables in the (finite) $A_4$ cluster algebra, both of these integrals are Laurent polynomials in terms of the initial cluster ${\bf x}$, 
so to deform them we can just take arbitrary linear combinations of the Laurent monomials that appear.

\begin{thm}\label{A4thm} 
The conditions 
\begin{equation} \label{condpar}
b_1=1=b_4 
\end{equation}
on the parameters $a_i$, $b_i$ (for $i=1,4$) in  (\ref{A4maps}) are necessary and sufficient 
for the first integrals defined by (\ref{A4ints}) in the periodic case to deform to a pair of rational 
conserved quantities  for  
the symplectic map $\varphi=\mu_4\mu_3\mu_2\mu_1$ that are in involution, i.e.\  they satisfy
(\ref{invol}) with respect to the 
Poisson bracket (\ref{bracket}). Hence 
the resulting two-parameter family of 
maps $\varphi$ is Liouville integrable, with  the 
two functionally independent commuting integrals 
\begin{align*}
I_1= &\frac{1}{ {x_1}  {x_2}  {x_3}  {x_4}}
\Big( {a_1}   {a_4}    {x_1}  {x_2}+ {a_1} 
 {a_4}^2  
 {x_1}  {x_2}  {x_3}+ {a_1} {x_1}  
 {x_2}  {x_3} 
 + {a_1}   {a_4} 
 {x_1}  {x_2}  {x_3}^2+ {a_1} 
 {a_4}    {x_1}  {x_4}  \\ 
& + {a_1}  
 {a_4}    {x_1}  {x_2}^2  {x_4} 
 + {a_1}  {a_4}  
 {x_3}  {x_4}+ {a_1}
 {a_4}   {x_1}^2  
 {x_3}  {x_4}+ {a_4}  
 {x_2}  {x_3}
 {x_4} 
 + {a_1}^2  {a_4}  {x_2}  {x_3}  {x_4}+  
 {a_4} 
 {x_1}^2  {x_2}  {x_3}  {x_4}  \\ 
& + {a_1}  
 {a_4}  {x_2}^2  {x_3}  {x_4} 
  + {a_1} 
 {a_4}    {x_1}  {x_3}^2  {x_4}+ 
 {a_1}  {a_4}   {x_1}  
 {x_2}  {x_4}^2+ {a_1}
 {x_1}  {x_2}  
 {x_3}  {x_4}^2\Big) , \\
I_2=&\frac{
 ( {a_1} +  {x_2}) (  {x_1}+  {x_3}) \left(  {a_4} 
 +  {x_3}\right) \left( {x_2}+   {x_4}\right) 
 \left(  {x_1}  {x_2}+
 {a_4}   {x_1}  {x_2}  {x_3}+   {x_1}  {x_4}+ 
 {x_3}  {x_4}+ {a_1}  {x_2}  {x_3}  {x_4}\right)}{ {x_1}  {x_2}^2
 {x_3}^2  {x_4}} .
 \end{align*}
\end{thm}
\begin{prf}
The calculation of the conditions on the coefficients of the monomials appearing in the deformed versions of the integrals (\ref{A4ints}) is direct, and leads to the above 
forms of $I_1,I_2$ together with the requirement that $b_1$ and $b_4$ should both equal 1. An explicit calculation of their Poisson bracket then shows that the deformed integrals 
are also in involution, as required for Liouville integrability. 
\end{prf}

\begin{table}[h!]
  \begin{center}
    \caption{Prime factors in an orbit of the integrable deformed $A_4$ map with $a_1=2, a_4=3$.}
    \label{factortable} 
\scalebox{0.8}{
    \begin{tabular}{ | r|| r| r| r| r |r|r|r|r|r|r|r|r|} %
\hline
      $n$ & 0 &1&2&3&4&5&6&7&8&9&10 & 11\\
\hline 
&&&&&&&&&&&&\\
$x_1$  & 1 & 3         &   3               &      3          & 7                           &
 $\tfrac{2^2}{7}$ &  $\tfrac{151}{2^2\cdot5}$ &
$\tfrac{5\cdot11\cdot61}{7\cdot151}$&
$\tfrac{7\cdot 251}{11\cdot61}$ &
$\tfrac{3\cdot11\cdot571}{5^2\cdot251}$&
$\tfrac{3\cdot5^2\cdot7\cdot5653}{11\cdot19\cdot23\cdot571}$ &
$\tfrac{3\cdot19\cdot 23\cdot54403}{7\cdot137\cdot5653}$ 
 \\ 
&&&&&&&&&&&&\\
$x_2$ & 1  & $2^2$ &  $2^2$         & $2\cdot 5$ &$\tfrac{3}{2}$&
$\tfrac{2\cdot 29}{5\cdot 7}$ & $\tfrac{643}{2^3\cdot7}$  &
$\tfrac{2^3\cdot 3\cdot23}{151}$ &
$\tfrac{5233}{5^2\cdot61}$&
 $\tfrac{2\cdot61613}{19\cdot23\cdot251}$ &
 $\tfrac{1031\cdot5519}{11\cdot137\cdot571}$ & 
$\tfrac{2\cdot11\cdot569\cdot42043}{5^2\cdot353\cdot5653}$
\\ 
&&&&&&&&&&&&\\
 $x_3$ & 1 &  5 &          13               &      2        & $\tfrac{13}{5}$ &
 $\tfrac{3^2\cdot 13}{7^2}$ &  $\tfrac{2\cdot71}{11}$ &
$\tfrac{11\cdot17\cdot89}{ 5^2\cdot151}$&
$\tfrac{79\cdot3529}{11\cdot19\cdot 23\cdot 61}$&
$\tfrac{1431173}{7\cdot137\cdot251}$&
$\tfrac{7\cdot73\cdot51539}{5^2\cdot353\cdot571}$ & 
$\tfrac{13\cdot17\cdot43\cdot237379}{7\cdot5653\cdot7507}$
 \\ 
&&&&&&&&&&&&\\
$x_4$ & 
1  & 
 $2^4$ &
$\tfrac{5}{2}$& 
$\tfrac{2\cdot 7}{5}$& 
$\tfrac{2\cdot 11}{7}$ &
$\tfrac{ 2^3\cdot 5^2}{7\cdot11}$&  
$\tfrac{7\cdot19\cdot 23}{2^3\cdot5^2}$ &
$\tfrac{2^6\cdot 7\cdot137}{19\cdot23\cdot151}$&
$\tfrac{2\cdot5^2\cdot 151\cdot353}{7\cdot11\cdot61\cdot137}$&
$\tfrac{2\cdot11\cdot61\cdot7507}{5^2\cdot251\cdot353}$&
 $\tfrac{19\cdot 101\cdot 251\cdot359}{11\cdot571\cdot7507}$ &
$\tfrac{2^8\cdot11\cdot571\cdot109943}{7\cdot19\cdot101\cdot359\cdot5653}$
 \\
&&&&&&&&&&&&\\
\hline 
    \end{tabular}
}
  \end{center}
\end{table}


To determine the singularity structure of the integrable map $\varphi$ we consider a particular rational orbit  with 
parameters $a_1=2,a_4=3$ and all initial $x_j$ equal to 1 (see Table 1). Applying the empirical $p$-adic method 
from \cite{hkq} once more, we observe that in the numerators of $x_2$ and $x_3$ there are certain primes that do not appear elsewhere, e.g.\ there are isolated values of $n$ where $|x_{2,n}|_p=p^{-1}$ for $p=29,643,5233,61613$,  
and similarly there are isolated $n$ where $|x_{3,n}|_p=p^{-1}$ for $p=17,71,79,89,3529, 1431173$. On the other hand, for $p=61,151,251,571$ 
there are particular values of $n$ where $|x_{1,n}|_p=|x_{2,n}|_p=|x_{3,n}|_p=|x_{4,n}|_p=p$ and also 
$|x_{1,n-1}|_p=p^{-1}$, $|x_{4,n+1}|_p=p^{-1}$. Also for $p=137,353,7507$ there is a pattern where $p$ first appears in the numerator of $x_4$, then in its denominator at the next step, then successively in the denominators 
of $x_3,x_2,x_1$, before appearing in the numerator of $x_1$, then disappearing at the  7th step (some of the 
factorizations required to see this are omitted from Table 1 for reasons of space); the product of primes $19\cdot 23$  
exhibits the same pattern, although these primes also appear separately elsewhere. 
These four singularity patterns in the iterates of $\varphi$ suggest introducing four tau functions $\eta_n,\theta_n,\si_n,\tau_n$, where the first two have weight two and the last two have weight one, such that 
\beq\label{A4tauf} 
\tilde{\pi}: \qquad
x_{1,n}=\frac{\si_n\tau_{n+1}}{\si_{n+1}\tau_n}, \quad 
x_{2,n}=\frac{\eta_n}{\si_{n+2}\tau_n}, \quad 
x_{3,n}=\frac{\theta_n}{\si_{n+3}\tau_n}, \quad 
x_{4,n}=\frac{\si_{n+5}\tau_{n-1}}{\si_{n+4}\tau_n}, \quad 
\eeq  
and direct substitution into the recurrence versions of (\ref{A4maps}) with $b_1=1=b_4$, replacing $x_j\to x_{j,n}$, $x_j'\to x_{j,n+1}$, gives 
the system 
\beq\label{A4system}\begin{array}{rcl}
\tau_{n+2}\si_n& =& \tau_n\si_{n+2}+a_1\, \eta_n, \\ 
\eta_{n+1}\eta_n &= &\si_{n+1}\tau_{n+2}\theta_n + \si_{n+2}\si_{n+3}\tau_n\tau_{n+1}, \\ 
\theta_{n+1}\theta_n & = & \si_{n+5}\tau_{n-1}\eta_{n+1}+\si_{n+3}\si_{n+4}\tau_n\tau_{n+1}, \\
\si_{n+6}\tau_{n-1}& = & \si_{n+4}\tau_{n+1}+a_4\, \theta_{n+1}. 
\end{array}
\eeq 
Initial evidence that this system has the Laurent property is provided by setting 
$\si_0=\cdots =\si_5=\eta_0=\theta_0=\tau_{-1}=\tau_0=\tau_1=1$,  corresponding to all initial $x_{j,0}=1$, 
$j=1,2,3,4$ as in Table 1, and iterating the above with $a_1=2$, $a_4=3$, which produces integer-valued 
tau functions as in Table 2.

\begin{table}[h!]
  \begin{center}
    \caption{Tau functions for the same orbit of the 
deformed $A_4$ map as in Table 1.}
    \label{factortable} 
\scalebox{0.7}{
    \begin{tabular}{ | r|| r| r| r| r |r|r|r|r|r|r|} 
\hline
      $n$ & 0 &1&2&3&4&5&6&7&8&9 \\ 
\hline 
&&&&&&&&&&  
\\
 $\tau_{n+1}$ & 1 & 3 &   9  &  27 & 189   &
 1728 & 97848 & 2608848 & 
64408608 & 
3516556032 
 \\ 
&&&&&&&&&& 
\\
$\eta_n$ & 1  & 4  &  12 & 90 &  648 &
37584 & 19999872 & 3399542784 & 
1546939772928 & 
1748502507552768 
\\ 
&&&&&&&&&& 
\\
 $\theta_n$ & 1 &  5 & 39 & 288 & 8424 & 
454896 &  212004864 &  74543597568 & 
59937513504768 & 
487379529497051136 
 \\ 
&&&&&&&&&& 
\\ 
$\si_{n+5}$ & 
1  & 16 & 120 & 1008 & 9504 &
 172800 &24164352 & 1272692736 & 
140540313600 & 
15780710449152 
 \\
&&&&&&&&&& 
\\
\hline 
    \end{tabular}
}
  \end{center}
\end{table}

\begin{figure}
 \centering
\label{quiverA4}
\epsfig{file=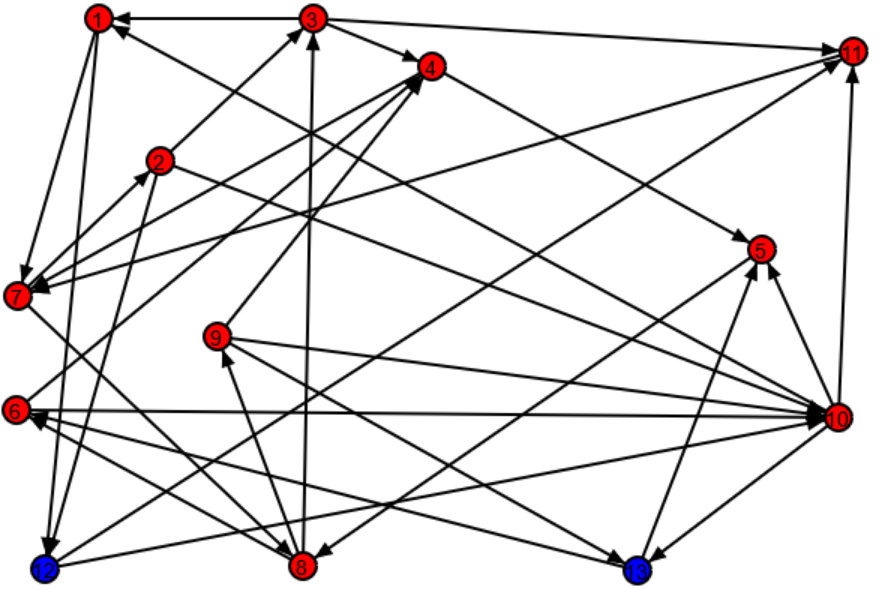, height=3in, width=3.5in}
\caption{The initial quiver associated with the extended exchange matrix (\ref{A4tauBt}).} 
\end{figure} 

If the initial data for (\ref{A4system}) is regarded as a cluster, that is  $$(\tx_1,\ldots,\tx_{11})=(\si_0,\ldots,\si_5,\eta_0,\theta_0,\tau_{-1},\tau_0,\tau_1),$$
then  
the pullback of the symplectic form (\ref{omegaA4}) under the map $\tilde{\pi}$ defined by (\ref{A4tauf}) is 
$$ 
\tilde{\om}=\tilde{\pi}^*\om =\sum_{i<j}b_{ij}^*\, \rd\log\tx_i\wedge\rd\log\tx_j, 
$$  
where $B^*=(b^*_{ij})$ is the exchange matrix 
\beq\label{A4tauB}
B^*=\left(\begin{array}{ccccccccccc}
0 & 0 & -1 & 0 & 0 & 0 & 1 & 0 & 0 & -1 & 0 \\ 
   & 0 &  1 & 0 & 0 & 0 & -1 & 0 & 0 & 1 & 0 \\ 
   &    &  0 & 1 & 0 & 0 & 0  & -1 & 0 & 0 & 1 \\ 
   &    &     & 0 & 1 &-1 & 1 &   0 &-1 & 0 & 0 \\
   &    &     &    & 0 & 0 & 0 & 1  &  0 &-1 & 0 \\ 
   &    &    &  &    & 0 & 0 & -1 &  0 &  1 & 0 \\
   &    &     &   &    &    & 0 & 1  & 0 &  0 & -1 \\
  &    &     & $*$   &   &     &    & 0  & 1 & 0 & 0   \\ 
   &   &     &    &   &     &    &    &  0 & 1 & 0  \\ 
  &   &      &    &   &     &    &     &    &  0 & 1 \\ 
  &   &      &     &  &     &     &    &    &     & 0 
\end{array}  
\right) 
\eeq 
(since the matrix is skew-symmetric, for brevity we put an asterisk to represent the terms below the diagonal). 
As in the $A_3$ case, this is sufficient to generate a sequence of mutations for the tau functions in the original undeformed system, but in order to include the parameters $a_1,a_4$ it it necessary to add these as frozen variables.  

\begin{thm}\label{LPA4taus}
The sequences of tau functions $(\tau_n)$,  $(\eta_n)$,  $(\theta_n)$, $(\si_n)$ for the integrable map 
$\varphi=\mu_4\mu_3\mu_2\mu_1$ defined by (\ref{A4maps}) with $b_1=b_4=1$ 
consist of elements of the Laurent polynomial ring 
$\Z_{>0}[a_1,a_4,
\si_{0}^{\pm 1},\si_{1}^{\pm 1},\si_{2}^{\pm 1},\si_{3}^{\pm 1},\si_{4}^{\pm 1},\si_{5}^{\pm 1},
\eta_{0}^{\pm 1},\theta_{0}^{\pm 1},\tau_{-1}^{\pm 1},\tau_{0}^{\pm 1},\tau_{1}^{\pm 1}]$,  being generated 
by a sequence of mutations in a cluster algebra defined by the exchange matrix (\ref{A4tauB}) 
with the addition of two frozen variables, corresponding to the quiver shown in Figure 4. 
\end{thm}

\begin{prf}
We take an extended cluster 
$$\tilde{\bf x}=(\tx_1,\ldots,\tx_{13})=(\si_0,\ldots,\si_5,\eta_0,\theta_0,\tau_{-1},\tau_0,\tau_1,a_1,a_4), $$ 
with the coefficients $a_1,a_4$ corresponding to additional frozen nodes in the quiver associated with 
$\tilde{B}^*=(b^*_{ij})$, the extended exchange matrix given by  
\beq\label{A4tauBt}
\tilde{B}^*=\left(\begin{array}{ccccccccccc}
0   & 0  & -1 & 0 & 0 & 0 & 1 & 0 & 0 & -1 & 0 \\ 
0   & 0  &  1 & 0 & 0 & 0 & -1 & 0 & 0 & 1 & 0 \\ 
1   & -1 &  0 & 1 & 0 & 0 & 0  & -1 & 0 & 0 & 1 \\ 
0   & 0  & -1 & 0 & 1 &-1 & 1 &   0 &-1 & 0 & 0 \\
0   & 0  &  0 & -1& 0 & 0 & 0 & 1  &  0 &-1 & 0 \\ 
0   & 0  &  0 & 1  & 0  & 0 & 0 & -1 &  0 &  1 & 0 \\
-1  & 1  &  0 &-1 &  0 & 0  & 0 & 1  & 0 &  0 & -1 \\
0  & 0  &  1  &  0 &-1  &  1 &-1  & 0  & 1 & 0 & 0   \\ 
0  & 0  &  0  &  1  & 0 & 0  & 0  &-1  &  0 & 1 & 0  \\ 
1  & -1 &  0  & 0 & 1 & -1 & 0 & 0  &-1  &  0 & 1 \\ 
0  & 0  & -1  & 0 & 0 & 0  & 1 & 0  & 0  & -1 & 0  \\
-1 &-1  &  0  & 0  & 0  &  0   & 0    & 0   & 0  &  1   & 1   \\ 
0  & 0  & 0   &  0 & 1 &  1   & 0  & 0 &-1 &-1 & 0
\end{array}  
\right)
\eeq 
(here we have shown the full matrix so that the exponents of all the exchange relations are visible in each 
column). The initial quiver is shown in Figure 4. Mutating at node 1 gives the exchange relation
$$ 
\tmu_1:\quad \tau_2\si_0=
\tau_0\si_{2}+a_1\, \eta_0, 
$$ 
producing the new cluster 
$\tmu_1(\tilde{\bf x})= (\tau_2,\si_1,\ldots,\si_5,\eta_0,\theta_0,\tau_{-1},\tau_0,\tau_1,a_1,a_4)$,
and subsequently applying mutations $\tmu_7,\tmu_8,\tmu_9$ successively generates exchange relations 
corresponding to the other three equations in (\ref{A4system}) for $n=0$, with the 
result being the cluster 
 $\tmu_9\tmu_8\tmu_7\tmu_1(\tilde{\bf x})= (\tau_2,\si_1,\ldots,\si_5,\eta_1,\theta_1,\si_{6},\tau_0,\tau_1,a_1,a_4)$. 
To generate each new instance of the four equations in (\ref{A4system}) with the index $n$ increased by 1, 
it is necessary to apply a similar block of four mutations.
Let us define the following composition of four mutations by 
$$ 
\hat{\mu}_{ij}
:=
\tmu_i\tmu_8\tmu_7\tmu_j, 
$$ 
and to index mutations we use $\overline{10},\overline{11}$ to distinguish nodes 10 and 11 from nodes with single-digit labels. 
Then if we take 
a particular composition 
of 36 mutations given by 9 of these blocks of four, namely 
$$
\hat{\hat{\mu}}  :=  \hat{\mu}_{6\overline{11}}\,\hat{\mu}_{5\overline{10}}\,\hat{\mu}_{49}\,
\hat{\mu}_{36}\,\hat{\mu}_{25}\, 
\hat{\mu}_{14}\,\hat{\mu}_{\overline{11}3}\,\hat{\mu}_{\overline{10}2}\,\hat{\mu}_{91} 
 =  \tmu_{687\overline{11}587\overline{10}4879387628751874\overline{11}873\overline{10}8729871}
$$
(where in the second expression the notation from (\ref{munotn}) has been reused), then 
the quiver returns to its starting position;  so we have 
$$ 
 \hat{\hat{\mu}}(\tilde{B}^*)= \tilde{B}^*, \qquad 
\hat{\hat{\mu}}(\tilde{\bf x})=(\si_9,\si_{10},\si_{11},\si_{12},\si_{13}, \si_{14},\eta_9,\theta_9, \tau_8,\tau_9,\tau_{10},a_1,a_4),
$$ 
with the index of each of the tau functions increased by 9. Thus by repeatedly applying these 9 blocks of four mutations, all 
of the tau functions for the integrable map are produced from clusters in the cluster algebra defined by  (\ref{A4tauBt}). 
\end{prf}

\section{Reductions of the discrete sine-Gordon equation}\label{sgredn}
\setcounter{equation}{0}

In this section we consider two examples  of four-dimensional maps that arise as reductions of the lattice sine-Gordon equation 
introduced in  \cite{Hir}, that is  
\begin{equation} \label{sinG}
a_1(x_{n,m} x_{n+1,m+1}-x_{n+1,m}x_{n,m+1})+a_2x_{n,m}x_{n+1,m}x_{n,m+1}x_{n+1,m+1}=a_3\;, 
\end{equation}
where $a_j$, $j=1,2,3$ are arbitrary parameters. Travelling waves of (\ref{sinG}) are obtained by imposing periodicity 
under shifts by $N$ steps in one lattice direction together with $M$ steps in the other direction, so that 
$$u_{n+N,m+M}=u_{n,m}\implies u_{n,m}=x_k,\quad k=Mn-Nm; $$
this is called the $(N,M)$ reduction. 

The two examples we consider below each correspond to particular orientations of the affine $A_3^{(1)}$ Dynkin diagram, as in Figure 5 
(where the notation $\tilde{A}_{p,q}$ means there are $p$ clockwise arrows and $q$ anticlockwise arrows).

\begin{figure}
\centering
\begin{subfigure}{.5\textwidth}
  \centering
  \includegraphics[width=.7\linewidth]{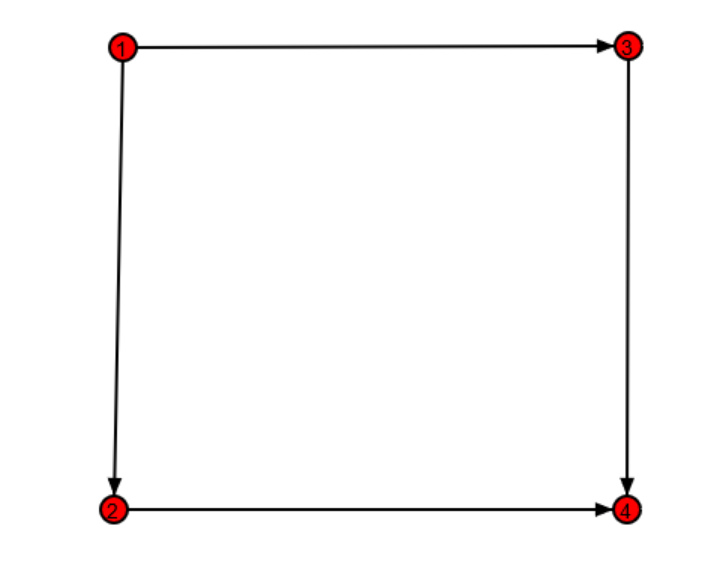}
  \caption{The quiver $\tilde{A}_{2,2}$.}
  \label{quiver2} 
\end{subfigure}%
\begin{subfigure}{.5\textwidth}
  \centering
  \includegraphics[width=.8\linewidth]{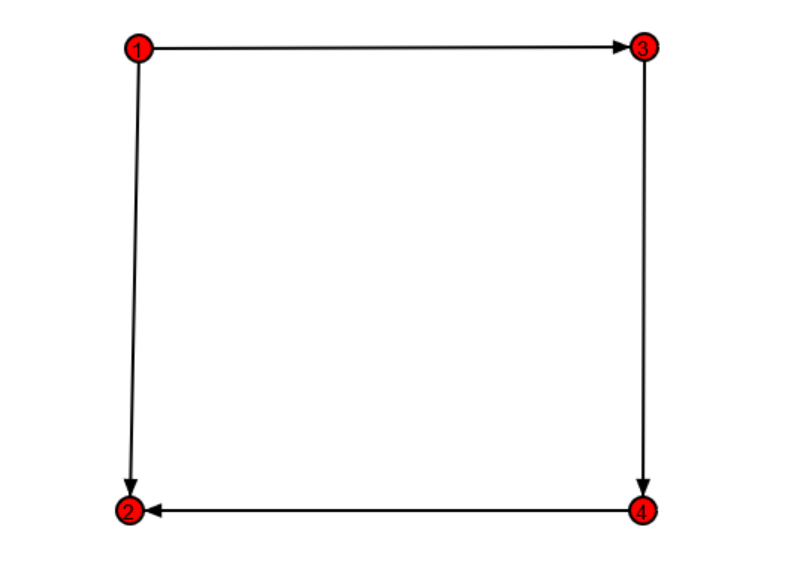}
  \caption{The quiver $\tilde{A}_{3,1}$.}
  \label{quiver3} 
\end{subfigure}
\caption{Two orientations of the $A_3^{(1)}$ Dynkin diagram.}
\label{affineA3}
\end{figure}

\subsection{(2,2)  periodic reduction of the lattice sine-Gordon equation} \label{exsing1}

Let us consider the quiver with exchange matrix 
\begin{equation*}
B= \left( \begin{array}{cccc} 0 & 1 & 0 & 1 \\
-1 & 0 & -1 & 0 \\
0 & 1 & 0 & 1 \\
-1 & 0 & -1 & 0
\end{array} \right);
\end{equation*}
this is mutation equivalent to $\tilde{A}_{2,2}$ as in Figure 5(a), which corresponds to the exchange matrix $\mu_3(B)$.
Then for $k=1,2,3,4$ we take   
 the function $$g_k(x)=\frac{a_1 x+a_3}{a_2 x+a_1},$$ for arbitrary parameters $a_1,a_2,a_3$, so that the 
exchange relation \eqref{clustmut} contains the function 
$$f_k(M^+_k,M^-_k)=M^+_k g_k\left(\frac{M^-_k }{M^+_k }\right)=M^+_k \frac{a_1 M^-_k +a_3 M^+_k}{a_2 M^-_k+a_1M^+_k }\;.$$ 
Next, we consider a sequence of mutations 
which leaves matrix $B$ 
invariant, specifically   
 $$\varphi (B,\mathbf{x}):={\mu}_{3}{\mu}_{1}{\mu}_{4} {\mu}_{2}(B,\mathbf{x})=(B,\tilde{\mathbf{x}}), 
 \ \text{where} \ 
\tilde{\mathbf{x}}=(\tilde{x}_1,\tilde{x}_2,\tilde{x}_3,\tilde{x}_4)$$ and
\begin{align*}
\tilde{x}_2&=\frac{1}{x_2}\left( \frac{a_1 x_1 x_3+a_3}{a_2 x_1 x_3+a_1} \right)\;, \ 
\tilde{x}_4=\frac{1}{x_4}\left( \frac{a_1 x_1 x_3+a_3}{a_2 x_1 x_3+a_1} \right)\;, \\ 
\tilde{x}_1&=\frac{1}{x_1}\left( \frac{a_1 \tilde{x}_2 \tilde{x}_4+a_3}{a_2 \tilde{x}_2 \tilde{x}_4+a_1} \right)\;, \ 
\tilde{x}_3=\frac{1}{x_3}\left( \frac{a_1 \tilde{x}_2 \tilde{x}_4+a_3}{a_2 \tilde{x}_2 \tilde{x}_4+a_1} \right)\;.
\end{align*}
So, according to Theorem \ref{thm2}, the map $\varphi:\mathbf{x}\mapsto\tilde{\mathbf{x}}$  
preserves the two form
\begin{equation*} 
\omega=\frac{1}{x_1 x_2} \rd x_1 \wedge \rd x_2+\frac{1}{x_1 x_4} \rd x_1 \wedge \rd x_4-\frac{1}{x_2 x_3} \rd x_2 \wedge \rd x_3
+\frac{1}{x_3 x_4} \rd x_3 \wedge \rd x_4\;.
\end{equation*}
In this case, the map $\varphi$ corresponds to the $(2,2)$ periodic reduction of the lattice sine-Gordon equation (\ref{sinG})
(see Figure 6).

\begin{figure}[h] 
\begin{center}
\begin{tikzpicture}[thin,scale=1.3, every node/.style={transform shape}]
\draw (0.0,1.0)-- (1.0,0.0);
\draw (1.0,0.0)-- (2.0,1.0);
\draw (2.0,1.0)-- (3.0,0.0);
\draw (3.0,0.0)-- (4.0,1.0);
\draw (4.0,1.0)-- (5.0,0.0);
\draw (0.0,1.0)-- (1.0,2.0);
\draw (1.0,2.0)-- (2.0,1.0);
\draw (2.0,1.0)-- (3.0,2.0);
\draw (3.0,2.0)-- (4.0,1.0);
\draw (5.0,0.0)-- (6.0,1.0);
\draw (4.0,1.0)-- (5.0,2.0);
\draw (5.0,2.0)-- (6.0,1.0);
\draw (1.0,2.0)-- (2.0,3.0);
\draw (2.0,3.0)-- (3.0,2.0);
\draw (3.0,2.0)-- (4.0,3.0);
\draw (4.0,3.0)-- (5.0,2.0);
\draw [->] (1.0,0.5) -- (1.0,1.5);
\draw [->] (2.0,1.5) -- (2.0,2.5);
\draw [->] (3.0,0.5) -- (3.0,1.5);
\draw [->] (4.0,1.5) -- (4.0,2.5);
\draw [fill=black] (0.0,1.0) circle (1pt);
\draw(-0.1,1.3) node {$x_1$};
\draw [fill=black] (1.0,0.0) circle (1pt);
\draw(1.1115929156639173,-0.21046306447702798) node {$x_2$};
\draw [fill=black] (2.0,1.0) circle (1pt);
\draw (2.0005444724900006,1.3) node {$x_3$};
\draw [fill=black] (3.0,0.0) circle (1pt);
\draw (3.1044801438134484,-0.21046306447702798) node {$x_4$};
\draw [fill=black] (4.0,1.0) circle (1pt);
\draw(4.008415815136897,1.3) node {$x_1$};
\draw [fill=black] (5.0,0.0) circle (1pt);
\draw (5.112351486460345,-0.2046306447702798) node {$x_2$};
\draw [fill=black] (1.0,2.0) circle (1pt);
\draw (0.91115929156639173,2.37) node {$x_2^{'}$};
\draw [fill=black] (3.0,2.0) circle (1pt);
\draw (2.981044801438134484,2.4) node {$x_4^{'}$};
\draw [fill=black] (6.0,1.0) circle (1pt);
\draw(6.101303043286428,1.3) node {$x_3$};
\draw [fill=black] (5.0,2.0) circle (1pt);
\draw (5.112351486460345,2.3) node {$x_2^{'}$};
\draw [fill=black] (2.0,3.0) circle (1pt);
\draw (2.1005444724900006,3.3) node {$x_3^{'}$};
\draw [fill=black] (4.0,3.0) circle (1pt);
\draw (4.108415815136897,3.3) node {$x_1^{'}$};
\end{tikzpicture}
\caption{The  $(2,2)$ staircase periodic reduction of the quadrilateral equation \eqref{sinG}}
\end{center}
\end{figure}
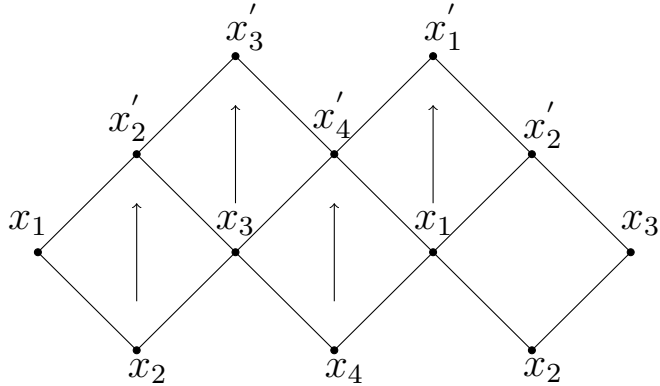
\label{red22}

The matrix $B$ (and hence $\om$) is degenerate, of rank two. To obtain a symplectic map, we take a pair of monomials corresponding to 
an integer basis for 
$$ 
\mathrm{im}\, B=<(1,0,1,0)^T,(0,1,0,1)^T>,
$$ 
namely 
$$ 
\pi: \qquad y_1=x_1x_3, \quad y_2=x_2x_4.
$$ 
Under the projection $\pi$ defined above, $\om$ is the pullback of the symplectic 
form 
$$ 
\hat{\om}=\frac{1}{y_1y_2}\,\rd y_1\wedge \rd y_2, 
$$
which is 
preserved by the induced map 
\beq 
\label{sGqrt} 
\hat{\varphi}: \quad 
\left(\begin{array}{c} y_1 \\ y_2 \end{array}\right) \mapsto 
\left( \begin{array}{c}\tilde{y}_1 \\ \tilde{y}_2 \end{array}
\right) , \quad \tilde{y}_2=\frac{1}{y_2}\left(\frac{a_1y_1+a_3}{a_2y_1+a_1}\right)^2, 
\tilde{y}_1=\frac{1}{y_1}\left(\frac{a_1\tilde{y}_2+a_3}{a_2\tilde{y}_2+a_1}\right)^2.
\eeq 
The above map has the first integral 
$$ 
K=\frac{a_2^2y_1^2y_2^2+2a_1a_2(y_1^2y_2+y_1y_2^2)+a_1^2(y_1^2+y_2^2)+2a_1a_3(y_1+y_2)+a_3^2}{y_1y_2}, 
$$ 
so it is Liouville integrable. In fact it is of QRT type: the level sets $K=\,$const are symmetric biquadratic curves, 
and $\hat{\varphi}=\iota_h \circ \iota_v =(\iota\circ\iota_v)^2$ where the involutions $\iota_h,\iota_v$ correspond to the horizontal and vertical switches 
on each level set, and $\iota:\,y_1\leftrightarrow y_2$.  For Laurentification of symmetric QRT maps, see \cite{hhkq}. 

In four dimensions, the other degrees of freedom in the original map $\varphi$ have essentially trivial  dynamics, since 
$$
\frac{\tx_1}{\tx_3}=\left(\frac{x_1}{x_3}\right)^{-1}, 
\qquad  
\frac{\tx_2}{\tx_4}=\left(\frac{x_2}{x_4}\right)^{-1}.
$$

\subsection{ (4,-1) periodic reduction of the lattice sine-Gordon equation } 
We consider the quiver with exchange matrix 
\begin{equation*}
B= \left( \begin{array}{rrrr} 0 & 1 & 0 & 1 \\
-1 & 0 & 1 & 0 \\
0 & -1 & 0 & 1 \\
-1 & 0 & -1 & 0
\end{array} \right).
\end{equation*}
The matrix $B$ is non-degenerate and satisfies $\mu_1(B)=\rho(B)$ for the cyclic permutation  
$\rho:(1,2,3,4)\mapsto(4,1,2,3)$, so it defines a cluster mutation-periodic quiver with period 1 \cite{FM}. 
Following the example in subsection \ref{exsing1}, we consider 
$$g_1(x)=x\,\left(\frac{a_1 +a_3 x}{a_2 +a_1x}\right)\;.$$
Here, $M^+_1={x_2 x_4}$, $M^-_1=1$ and 
$$f_1(M^+_1,M^-_1)=M^+_1 g_1\left(\frac{M^-_1 }{M^+_1 }\right)= 
\frac{a_1 x_2 x_4+a_3}{a_2 x_2 x_4+a_1} 
.$$ 
Hence, the appropriate analogue of Theorem \ref{thm2} (see Remark \ref{permperiodic}) implies that the map $\varphi=\rho^{-1}\mu_1$ given by 
\begin{equation} \label{41red}
\varphi:\, (x_1, x_2,x_3,x_4) \mapsto \left(x_2, x_3,x_{4},\frac{1}{x_1}\Big( \frac{a_1 x_2 x_4+a_3}{a_2 x_2 x_4+a_1} \Big) \right)
\end{equation}
preserves the symplectic form 
\begin{equation*} 
\omega=\frac{1}{x_1 x_2} \rd x_1 \wedge \rd x_2+\frac{1}{x_1 x_4} \rd x_1 \wedge \rd x_4 +\frac{1}{x_2 x_3} \rd x_2 \wedge \rd x_3 
+\frac{1}{x_3 x_4} \rd x_3 \wedge \rd x_4\;.
\end{equation*}
The map \eqref{41red} is associated with the $(4,-1)$ periodic reduction of the lattice  sine-Gordon equation \eqref{sinG}, 
and can be rewritten in recurrence form as  
\begin{equation*} 
a_1(x_{n} x_{n+4}-x_{n+1}x_{n+3})+a_2x_{n}x_{n+1}x_{n+3}x_{n+4}=a_3\;. 
\end{equation*}
Closed-form expressions for integrals of periodic reductions of the sine-Gordon equation were presented in \cite{KampRQ}
and their involutivity was proved in \cite{TranKQ}.

\section{Concluding remarks} 
\setcounter{equation}{0}

We have considered autonomous recurrences or maps obtained by including additional constant parameters in sequences of cluster mutations 
that generate completely periodic dynamics, and have shown that it is possible to preserve the presymplectic structure defined by the 
exchange matrix, and also (by imposing suitable constraints on the parameters) obtain Liouville integrable maps. Our starting point 
for showing Liouville integrability has been the fact that the original periodic maps admit first integrals defined by cyclic symmetric functions 
of variables along a period of the orbit. Only the examples of  $A_2$, $A_3$ and $A_4$ have been  dealt with here, but it would be instructive to make 
a more systematic study of such functions from the viewpoint of 
the associated  Poisson algebra in order to extend these results to cluster algebras defined by other finite type Dynkin diagrams. We have also treated more 
general types of mutations, involving M\"obius transformations, and shown that for some particular affine type exchange matrices these lead to reductions 
of the discrete sine-Gordon equation.  

The parameters $a_k,b_k$ appearing  in our deformed mutations have been assumed constant, but 
Theorem \ref{thm2} applies equally well to non-autonomous recurrences/maps. In 
particular, taking 
$$ 
a_k = \frac{y_k}{1+y_k}, \quad b_k=\frac{1}{1+y_k}
$$ 
in (\ref{fk}) 
leads to the expression for a mutation $\mu_k$ in a cluster algebra with coefficients \cite{fziv}, 
which themselves mutate according to 
$$ 
y_j'=\begin{cases}
y_k^{-1} & \text{if} \,\, j=k, \\
y_j \left(1+y_k^{-\text{sgn}(b_{jk})} \right)^{-b_{jk}}  & \text{otherwise}.
\end{cases}
$$
The dynamics of the coefficients generates the associated Y-system \cite{kun}. In \cite{hi}, 
it is shown that non-autonomous dynamics also arises from autonomous Y-systems in the case 
where the exchange matrix is degenerate: one of the simplest examples is provided by the Y-system 
$$ 
y_{n+7}y_n=
\frac{
(1+y_{n+6})(1+y_{n+1})}{(1+y_{n+4}^{-1})(1+y_{n+3}^{-1})}
$$ 
corresponding to the Somos-7 recurrence (\ref{s7}), solved in terms of the 
q-Painlev\'e V equation 
\beq \label{qpv}
x_{n+2}x_n =x_{n+1}+\al_n\,\mathfrak{q}^n, \qquad \al_{n+6}=\al_n, 
\eeq 
which is  a 
non-autonomous version of the Lyness recurrence. The fact that the period of $\al_n$ is 6 is important, since 
if $\mathfrak{q}=1$ and $\al_n$ is periodic with a period that is not a divisor of 6, then (\ref{qpv}) appears to exhibit 
chaotic dynamics \cite{cima}.

As another example based on the $A_2$ exchange matrix, taking $g_1(x)=\frac{a x+b}{c x+d}$ 
and letting the coefficients $a,b,c,d$ depend on the index $n$ gives the sequence of symplectic maps
$$ \varphi_n (x,y)=\left(y,\frac{a_n y+b_n}{x(c_n y+d_n)}\right) $$   
that corresponds to the non-autonomous nonlinear recurrence 
\begin{equation*}
x_{n+2}=\frac{a_n x_{n+1}+b_n}{x_n(c_n x_{n+1}+d_n)}\;.
\end{equation*}
Invariants of this recurrence when the coefficients are periodic were presented in \cite{Ladas} and have also been studied in the framework of 
QRT (and non-QRT) maps  with periodic coefficients \cite{Ramani,Roberts}.

\noindent \textbf{Acknowledgments:} This research was supported by 
Fellowship EP/M004333/1  from the Engineering \& Physical Sciences Research Council, UK, 
and grant 
 IEC\textbackslash R3\textbackslash 193024 from the Royal Society. 
All of the pictures of quivers were produced using Bernhard Keller's JavaScript mutation applet \cite{keller}. 
On behalf of all authors, the corresponding author states that there is no conflict of interest.


\begin{thebibliography}{99}

\bibitem{blanc} J. Blanc, Symplectic birational transformations of the plane, 
Osaka J. Math. 50 (2013) 573--590.

\bibitem{cima} A. Cima, A. Gasull and V. Ma\~nosa, 
Integrability and non-integrability of periodic non-autonomous Lyness recurrences, 
Dyn. Syst. 285 (2013) 18--38.

\bibitem{coxeter} H. Coxeter, Frieze patterns, 
Acta Arithmetica 18 (1971) 297--310.


\bibitem{dl} 
A. Doliwa and R. Lin, 
Discrete KP equation with self-consistent sources,
Phys. Lett. A 378 (2014) 1925--1931.

\bibitem{duistermaat} J.J. Duistermaat, 
Discrete Integrable Systems: QRT Maps and Elliptic Surfaces, 
Springer Monographs in Mathematics, vol. 304. Springer, 2010. 

\bibitem{eschr}J. Esch and T.D. Rogers, 
The screensaver map: dynamics on elliptic curves arises from polygonal folding, 
Discrete Comput. Geom. 25 (2001) 477--502.

\bibitem{eqr} C.A. Evripidou, G.R.W. Quispel and J.A.G. Roberts, 
Poisson structures for difference equations, 
 J. Phys. A: Math. Theor. 51 (2018) 475201.


\bibitem{Ladas}
J. Feuer, E.J. Janowski and G. Ladas,  Invariants for some rational recursive sequences with periodic coefficients, 
J. Difference Equ. Appl. 
2 (1996) 167--174.

\bibitem{fg} V.V. Fock and A.B. Goncharov,
Cluster ensembles, quantization and the dilogarithm, 
Ann. Sci. \'{E}c. Norm. Sup\'{e}r. 
42 (2009) 865--930.  

\bibitem{fz2} S. Fomin and A. Zelevinsky, 
 Y-systems and generalized associahedra, Ann. Math. 158 (2003) 977--1018.
 
\bibitem{fziv} S. Fomin and A. Zelevinsky, 
Cluster algebras IV: coefficients, Comp. Math. 143 (2007) 112--164. 

\bibitem{FH} 
A.P. Fordy and A.N.W. Hone, Discrete integrable systems and Poisson algebras from cluster maps, 
Comm. Math. Phys. 
325 (2014) 527--584.


\bibitem{FM} 
A.P. Fordy and R.J. Marsh, Cluster Mutation-Periodic Quivers and Associated Laurent Sequences, 
J. Algebr. Comb. 
34 (2011) 19--66.

\bibitem{gal} P. Galashin and P. Pylyavskyy, 
Quivers with subadditive labelings: classification and integrability, 
Math. Z. 295 (2020) 945--999.

\bibitem{gsv} M. Gekhtman, M. Shapiro and A. Vainshtein, 
Cluster algebras and Weil-Petersson forms, Duke Math. J. 127 (2005) 291--311.

\bibitem{ghkk} 
M. Gross,  P. Hacking,  S. Keel  and  M. Kontsevich,  
Canonical  bases  for cluster algebras, 
J. Amer. Math. Soc. 31 (2018) 497--608. 



\bibitem{hhkq} 
K. Hamad, A.N.W. Hone, P.H. van der Kamp and  G.R.W. Quispel, 
QRT maps and related Laurent systems, 
Adv. Appl. Math. (2018) 216--248.

\bibitem{Hir} 
R. Hirota,  Nonlinear partial difference equations III: Discrete Sine-Gordon equation, 
J. Phys. Soc. Jpn. 
43 (1977) 2079--2086.  

\bibitem{HirTs} 
R. Hirota and S. Tsujimoto, Conserved quantities of a class of nonlinear difference-difference equations, J. Phys. Soc. Jpn. 64 (1995) 3125--3127.

\bibitem{hkq} A.N.W. Hone, T.E. Kouloukas and  G.R.W. Quispel, 
Some integrable maps and their Hirota bilinear forms, J. Phys. A: Math. Theor. 51 (2018) 044004.

\bibitem{hi} A.N.W. Hone and R. Inoue, 
Discrete Painlev\'e equations from Y-systems, 
J. Phys. A: Math. Theor. 47 (2014) 474007.

\bibitem{in} R. Inoue and T. Nakanishi, 
 Difference equations and cluster algebras I: Poisson bracket for integrable difference equations, 
RIMS Kokyuroku Bessatsu 
B 28 (2011) 63--88. 



\bibitem{KampRQ}
 P.H. van der Kamp, O. Rojas and G.R.W. Quispel, Closed-form expressions for integrals of MKdV and sine-Gordon maps, 
J. Phys. A: Math. Theor. 
40 (2007) 12789. 

\bibitem{kanki}
M. Kanki, J. Mada,  K.M. Tamizhmani, T. Tokihiro,
Discrete Painlev\'e II equation over finite fields,
J. Phys. A: Math. Theor. 45 (2012) 342001.

\bibitem{keller} B.~Keller, Quiver mutation in JavaScript and Java/Mutation des carquois en JavaScript et Java, 
\url{https://webusers.imj-prg.fr/~bernhard.keller/quivermutation/}

\bibitem{kun} A. Kuniba, T. Nakanishi and J.  Suzuki, 
T-systems and Y-systems in integrable systems, J. Phys. A: Math. Theor. 44 (2011) 103001.

\bibitem{ls} K. Lee and R. Schiffler, Positivity for cluster algebras, Ann. Math. 182 (2015) 73--125.

\bibitem{lyness} R.C. Lyness, Note 1581, Math. Gaz. 26 (1942) 62.   


\bibitem{nak} T. Nakanishi, Periodicities in cluster algebras and dilogarithm identities, 
Representations of Algebras and Related Topics (EMS Series of Congress Reports), ed. A. Skowronski and K. Yamagata (Zurich: European Mathematical Society), pp 407--43, 2011.

\bibitem{pyl} P. Pylyavskyy, Zamolodchikov integrability via rings of invariants, 
J. Integrable Systems 1 (2016) xyw010. 

\bibitem{qrt} G.R.W. Quispel, J.A.G. Roberts and C.J. Thompson,
Integrable mappings and soliton equations, 
Phys. Lett. A 126 (1988) 
419--421.

\bibitem{Ramani} A. Ramani, B. Grammaticos and R. Willox,  Generalized QRT mappings with periodic
coefficients, 
Nonlinearity 
24 (2011) 113. 

\bibitem{Roberts}
J.A.G.  Roberts and D. Jogia, Birational maps that send biquadratic curves to biquadratic curves, 
J. Phys. A: Math. Theor. 
48  (2015) 08FT02.

\bibitem{tran} D.T. Tran, P.H. van der Kamp and G.R.W. Quispel, Sufficient number of integrals for the pth-order Lyness equation,
J. Phys. A: Math. Theor. 43 (2010) 302001.
	
\bibitem{TranKQ}
D.T. Tran, P.H. van der Kamp and G.R.W. Quispel, Involutivity of integrals of sine-Gordon, modified KdV and potential KdV maps, 
J. Phys. A: Math. Theor. 
44 (2011) 295206.

\bibitem{zabrodin} A.V. Zabrodin, A survey of Hirota's difference equations, 
Theoret. Math. Phys. 
113 
(1997) 1347--1392.

\bibitem{zam} Al.B. Zamolodchikov, 
On the thermodynamic Bethe ansatz equations for reflectionless ADE scattering theories,
Phys. Lett. B 253 (1991) 391--394.

\end{thebibliography}
\end{document}